\documentclass[11pt]{article}
\usepackage{geometry}                
\usepackage{multirow}  
 \usepackage{rotating}  
\usepackage{graphicx}
\usepackage{amssymb}
\title{Citations and impact of Dutch astronomy}
\author{P. Kamphuis and P.C. van der Kruit}

\begin{document}
\maketitle
\section{Introduction}
\hspace*{0.5 cm}The aim of this study is to make a bibliometric comparison of the
performance of research astronomers in the {\it Netherlands Research
School for Astronomy} (NOVA) with astronomers elsewhere. This is
complementary to similar studies as undertaken by the 
{\it Center for Science and Technology Studies} (CWTS), 
but has the specific feature of using
the {\it NASA Astrophysics Data System} (ADS)\footnote{The American and
European websites of ADS are {\tt http://adsabs.harvard.edu} and
{\tt http://esoads.}\linebreak[4]
{\tt eso.org}.} and possibilities offered by it. We will use various
indices for bibliometric performance for a sample of Dutch (NOVA)
astronomers to compare to samples of astronomers worldwide, from the
United States in general and from American top-institutes.

Secondly, we will consider the 
results of the {\it `Nederlands Observatorium van Wetenschap en
Technologie'} (NOWT), 
the Netherlands Observatory of Science and Technology, which regularly
publishes
a report {\it `Science and Technology Indicators'}\footnote{This study is
commissioned by the
{\it Ministery of Education, Culture and Science} (OCW) and performed jointly
by the
{\it Center for Science and Technology Studies} (CWTS) of Leiden University
and
the {\it Maastricht Economic and social Research and training centre on
Innovation and Technoly} (MERIT) of Maastricht University.}. We will
try and reproduce those results using publication lists from
institutions in the Netherlands, again using ADS, and examine and
discuss the conclusions and indications from these reports. 

\section{NASA Abstract Database Service}
\hspace*{0.5 cm}We investigate here first the reliability 
of the NASA {\it Astrophysics Data System} (ADS) for citation analysis
in order to address questions about bibliometric performance. 
Briefly, the ADS is a database for astronomical publications; it
provides scanned versions of articles in astronomy and astrophysics 
in almost all relevant journals, going back to the earliest volumes (in
relevant cases well into the nineteenth century), and of papers in proceedings
and observatory annals. For papers published since journals started electronic
subscriptions the user is linked to the journal sites. Also ADS provides for each
article a list of papers that have been cited and of papers that cite
that particular publication. In principle ADS is a complementary database
to the {\it Science Citation Index} used for NOWT. Also it can
easily order papers of a particular author or group of authors 
by the number of citations and normalised counts or normalised
citations\footnote{Normalised counts/citations are numbers of 
publications/citations where
each paper is counted with a weight that is the inverse of the number
of authors.}, etc. The ADS is a unique facility, available free of charge,
made available and maintained by NASA with public funds.

There have been a few investigations into the use of ADS
for bibliometric studies. Helmut Abt\footnote{A Comparison of the
citation counts in the Science Citation Index and the Nasa Astronomical
Data System (2004); {\tt
http://www.garfield.library.upenn.edu/papers/helmutabtorgstratastronv6y2004.html}.}
concludes that the correspondence is good. His abstract reads: {\it 
``From a comparison of 1000+ references to 20 papers in four fields of
astronomy (solar, stellar, nebular, galaxy), we found that the citation
counts in Science Citation Index (SCI) and Astronomical Data System
(ADS) agree for 85\%\ of the citations. ADS gives 15\%\ more citation counts
than SCI. SCI has more citations among physics and chemistry journals,
while ADS includes more from conferences. Each one misses less than 1\%\
of the citations.''} More specific to the determination of impact ratios, 
van der Kruit\footnote{Citation analysis
and the NASA Astrophysics Data System (2005), 
{\tt http://www.astro.rug.nl/\linebreak[4]
$\sim$vdkruit/jea3/homepage/ads.pdf}.} found that
studies based on ADS reproduce the results of studies by the CWTS very well
and ADS reliably provides information to perform bibliometric studies.

\section{The performance of NOVA astronomers}
\hspace*{0.5 cm}We use the ADS to investigate the distribution of various 
publication and
citation scores of NOVA astronomers. To this end we used a sample
comprising all staff members of the NOVA institutes in the 2003 -- 2009
periodand three comparison
samples. A list of NOVA researchers
was provided by Wilfried Boland, executive director 
of NOVA; this is the same list as has been used by the CWTS in its recent
bibliometric study of NOVA. We will designate this list as the `NOVA-list'.
We also used a list of `key-researchers' of
NOVA, which is of course a selection from the NOVA list itself.

The first random comparison sample has been taken from the membership 
directory of the International Astronomical Union 
(IAU)\footnote{Starting with www.iau.org/administration/membership/individual/186;
membership numbers run from 186 to 9605.}. 
Only `active members' were considered. After every selected member his/her
membership number $n$ was 
updated by adding 25. If no name was found the next number was taken, etc. 
This resulted in a list of order 400 astronomers. 
We then ignored names that had major difficulties in unique
identification in the ADS, such as `Li' among Chinese astronomers; this applied to 
about half the sample we had selected.  
The procedure was repeated for the American Astronomical Society 
(AAS; taking every 50th 
member)\footnote{Starting at
members.aas.org/directory/public\_directory/member\_details.cfm?ID=10000; 
AAS membership numbers run from 10,000 to 36,000.}
to produce the second comparison sample. Here we only selected `full members'.
As a final sample a sub-selection of the AAS was constructed by considering only the
faculty of the top 15 institutes of the US. These institutes follow from a
study by A.L. Kinney\footnote{The institutes are listed in Appendix A.}.
Note that this selection uses citation scores and therefore this sample contains
astronomers with high citation scores that often work in areas of astrophysics
with high citation rates. 
The samples comprised  79 (NOVA), 177 (TopUS), 172 (AAS)  and 193 (IAU) 
astronomers (the list of key-researchers has 26 persons).

\begin{table}[!tb]
\centering
{\small
\begin{tabular}{lccccc}
\hline
 & NOVA &TopUS & AAS & IAU &NOVA-kr\\
\hline
Number of people in sample & 79 &  177\ \  &172\ \  &193\ \ & 26\\
Number of papers & 90 & 94 & 26 & 58 &123\ \   \\
Number of first-author papers & 21 & 22& \ \ 8 & 19 & 23\\
Normalised number of papers & 26 & 25 & \ \ 7 & 20& 33\\
Number of normalised first-author papers & 11 & 10& \ \ 3 & \ \ 9& 13 \\
Number of citations & 3325\ \ \  & 4175\ \ \ & 544\ \  & 1042\ \ \  & 4558\ \ \ \\
Number of first-author citation & 795\ \ & 971 & 112 & 256& 1166\ \ \ \\
Number of normalised citations & 704\ \  & 929\ \  & 86 & 271\ \ &1213\ \ \  \\
$h$-index & 31 & 34 & 12 & 17& 39 \\
First-author $h$-index & 13 & 14 & \ \ 4 & \ \ 8& 14\\
Normalised $h$-index & 14 & 16 & \ \ 5 & \ \ 9 &16\\
First-author normalised $h$-index & \ \ 9 & 10 & \ \ 3 & \ \ 6&10 \\
Citations per paper & 36 & 43 & 21 & 18 & 38\\
Citations per first-author paper & 39 & 39 & 13 & 15& 45 \\
Normalised citations per normalised paper & 33 & 35 & 17 & 14&37 \\
Papers per year & \ \ \ 3.9 & \ \ \ 3.3 & \ \ \ 1.6 & \ \ \ 1.9 & \ \ \ 5.6\\
Citations per year & 131\ \  & 141\ \  & 34 & 34& 229\ \ \\
Normalised papers per year & \ \ \ \ 1.1 & \ \ \ \ 0.9 & \ \ \ \ 0.4 & \ \ \ \ 0.7& \ \ \ 1.5\\
Normalised citations per year & 34 & 33 & \ \ 6 & 10 &57\\
First-author papers per year & \ \ \ \ 0.9 & \ \ \ \ 0.9 & \ \ \ \ 0.5 & \ \ \ \ 0.6 &\ \ \ \ 1.2  \\
First-author citations per year & 40 & 33 & \ \ 8 & \ \ 9 & 59 \\
Number of publishing years & 25 & 30 & 18 & 30& 22\\
\hline
\end{tabular}
}
\caption{\small  Medians of the various distributions of publication and
citation scores for the samples of NOVA, topUS (15 top US institutions) 
tenured staff , AAS and IAU members and NOVA `key-researchers. 
The distributions (except for NOVA-kr)
are shown as histograms in Figures \ref{fpap}-\ref{fyearpubl}.}
\label{tabmed}
\end{table}%

>From the ADS we found for each person the number of publications (in a
refereed journal),
first-author publications, 
normalised\footnote{This normalisation is done by dividing the score for each paper 
by the number of authors. For example, a paper with five authors would 
contribute 0.2 to the normalised number of papers of each author.
The same is done for normalised citations.} publications and first-author
papers, and to all these sets of publications a sorted list of the number of
citations and normalised citations. We also noted the year of the first
(refereed) publication. From this we calculated various properties such as
(normalised) citations per paper, Hirsch-index\footnote{J.E. Hirsch,
Proc. Nat. Acad. Sci. 102(46), 16569 (2005); the value $h$ of the
$h$-index is defined such that the person involved is a (co-)author of $h$ papers
that have been cited more than $h$ times.}, (normalised) citations
and (normalised) publication per paper and per year (since that of the first
publication). 

It is well-known that the $h$-index of a person increases
with time, and for comparison between individuals therefore sometimes the
index is divided by the number of years that the person concerned has
been active in research. This assumes (usually incorrectly) that one's 
$h$-index increases linearly with time. 
A more serious shortcoming is that it is not
always calibrated between disciplines or (sub-)fields of scientific
research, where very different publication and citation cultures may
prevail. We will in this report use samples that give an idea of what a
typical $h$-index is for an astronomical researcher. Another
important effect is that of the number of authors on a paper. E.g. persons
contributing to highly cited papers presenting a new instrument or
survey all get the credit of an additional point on their 
$h$-index.\footnote{E.g., the Sloan Digital Sky Survey data release papers have
well over a hundred authors and receive hundreds of citations. Such a
paper will contribute one extra point to the $h$-index of all authors.}
Therefore we also calculated $h$-indices using normalised citations.
Furthermore, we also produced statistics using only papers on which the
researchers involved are first author.
\bigskip

\begin{figure}[!htb]
\centering
\includegraphics[width=14.5cm]{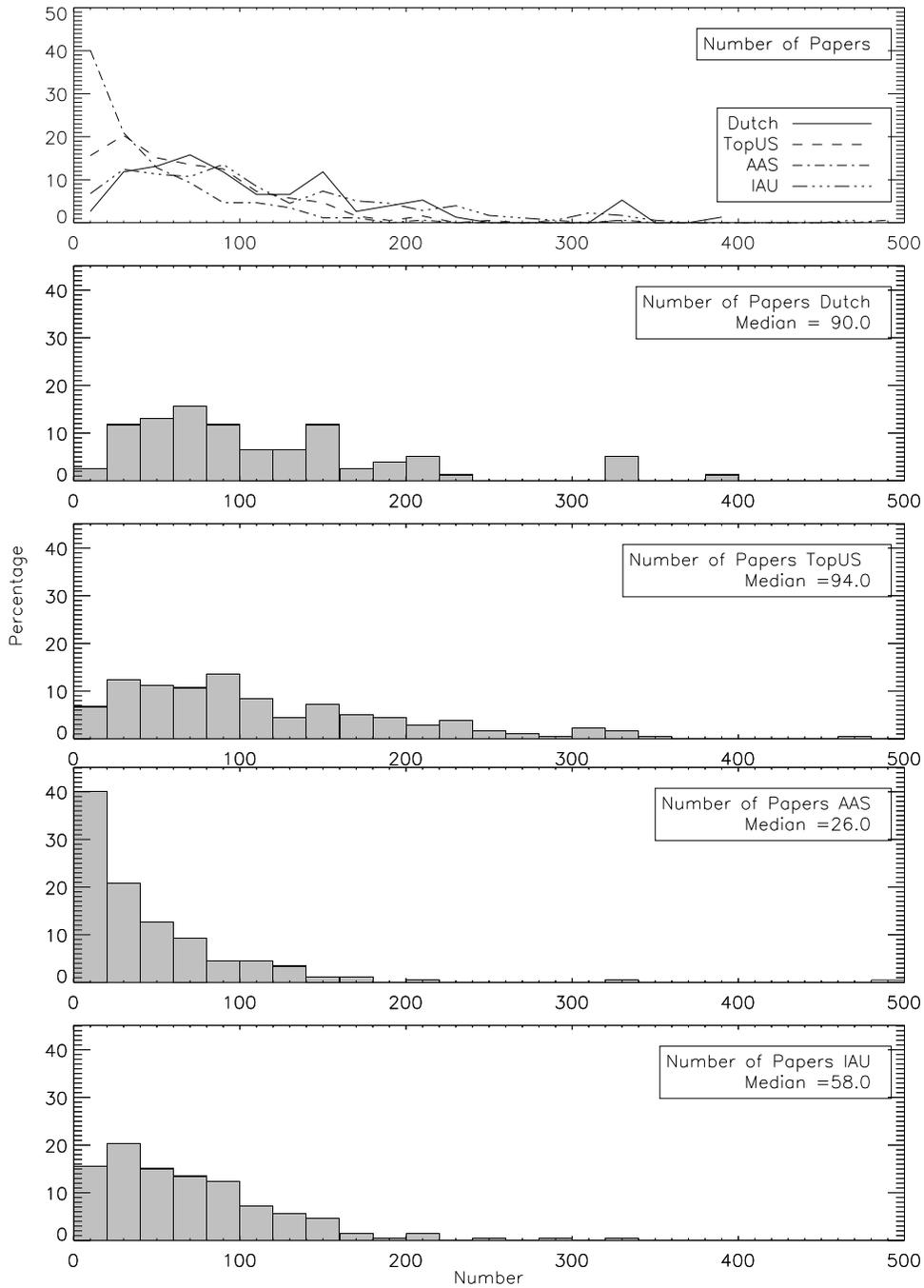}
\caption{\small  Histograms and a plot of the percentage of authors (y-axis) that has a 
certain number of papers (x-axis).  From top to bottom: 
(1) The Dutch (solid line) 
,the TopUS (dashed line), random AAS (dot-dashed line) and random IAU 
(three-dotted-dashed line) samples. 
(2) Histogram of the distribution of the Dutch sample.
(2) Histogram of the distribution of the TopUS sample. 
(3) Histogram of the distribution of the AAS sample. 
(4) Histogram of the distribution of the IAU sample.}
\label{fpap}
\end{figure}

\begin{figure}[htbp]
\centering
\includegraphics[width=14.5cm]{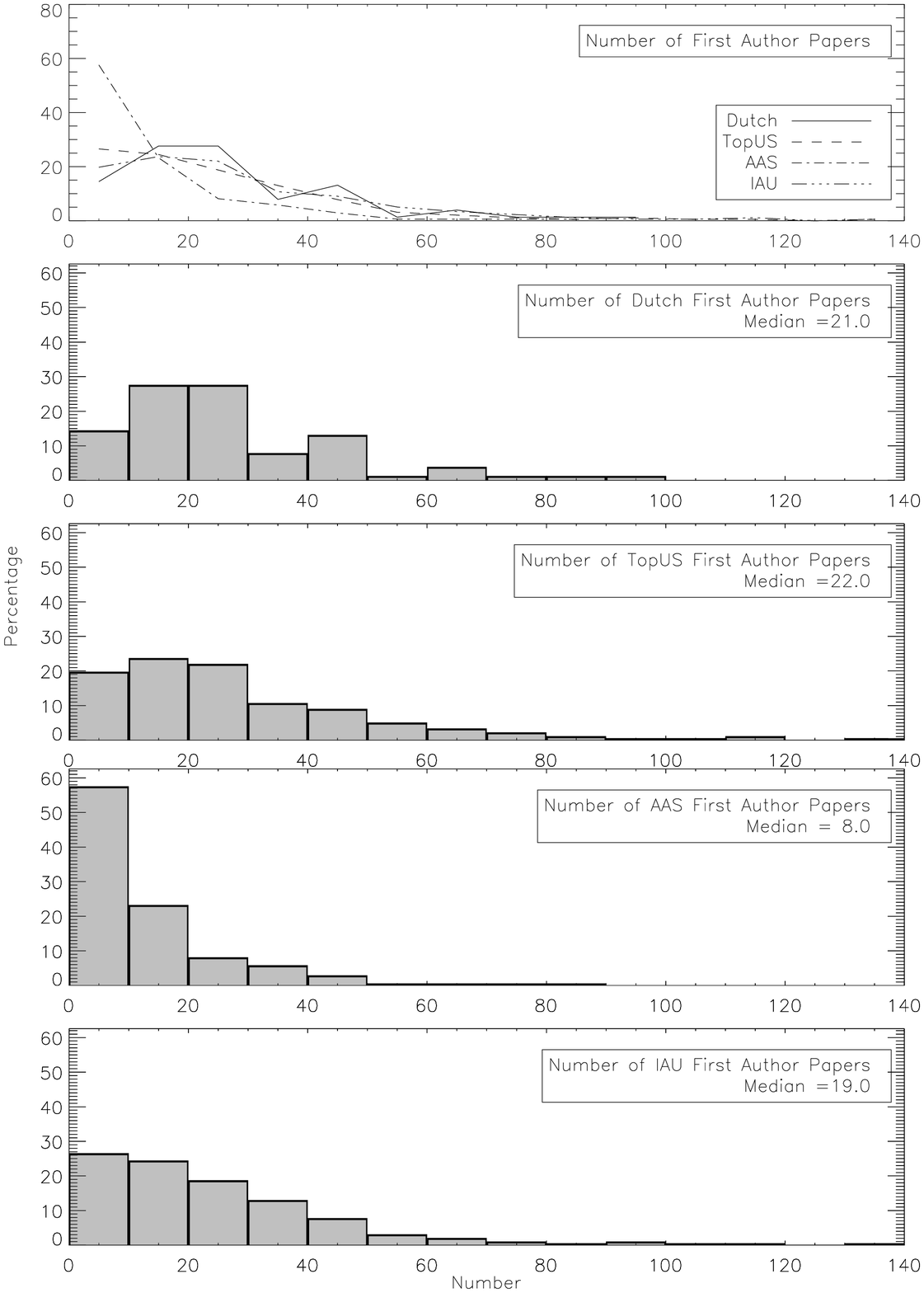}
\caption{\small As Figure \ref{fpap} but now only first author papers are considered.}
\label{fpapf}
\end{figure}

\begin{figure}[htbp]
\centering
\includegraphics[width=14.5cm]{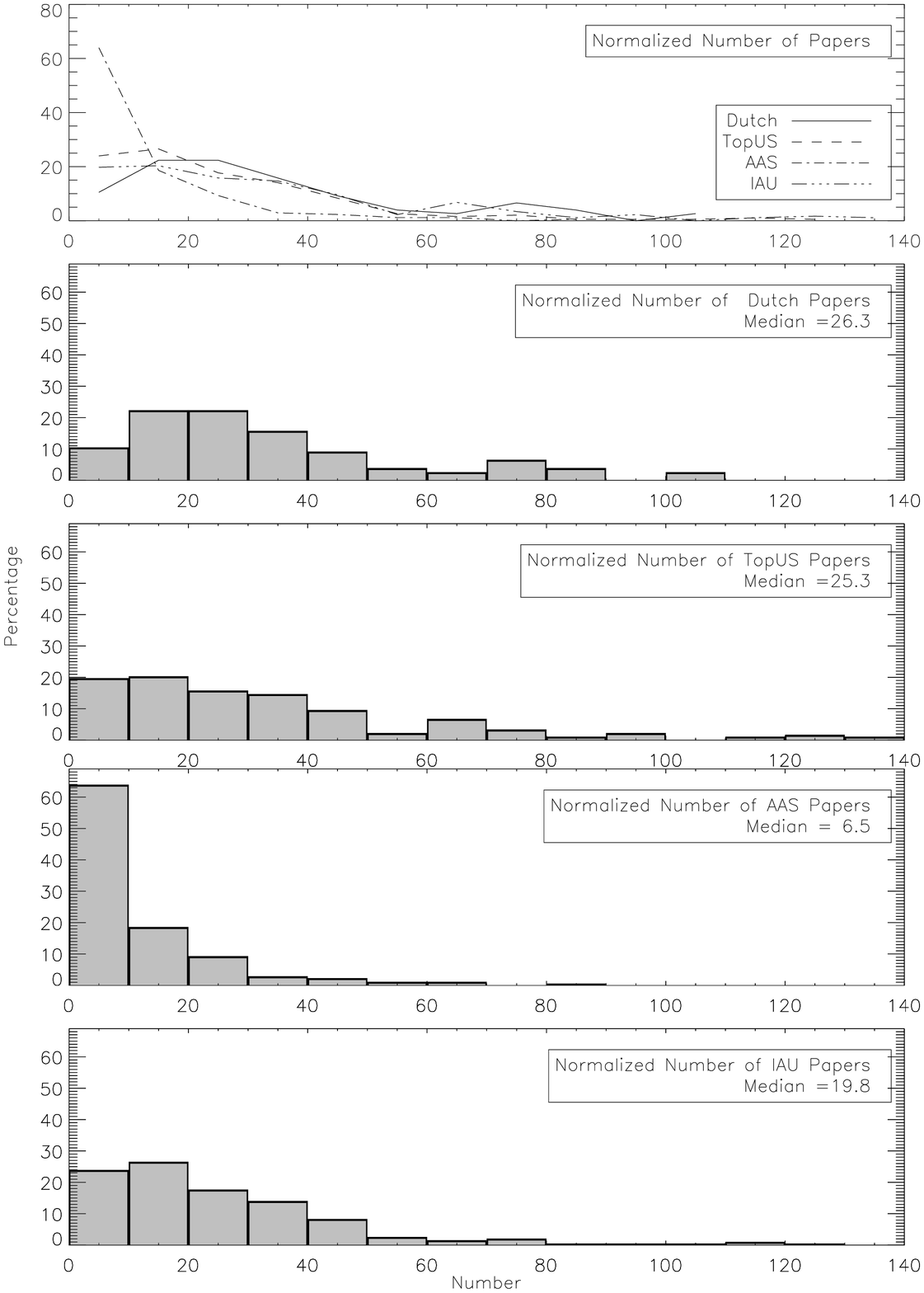}
\caption{\small As Figure \ref{fpap} but now the papers are normalised to the number of authors.}
\label{fpapnorm}
\end{figure}

\begin{figure}[htbp]
\centering
\includegraphics[width=14.5cm]{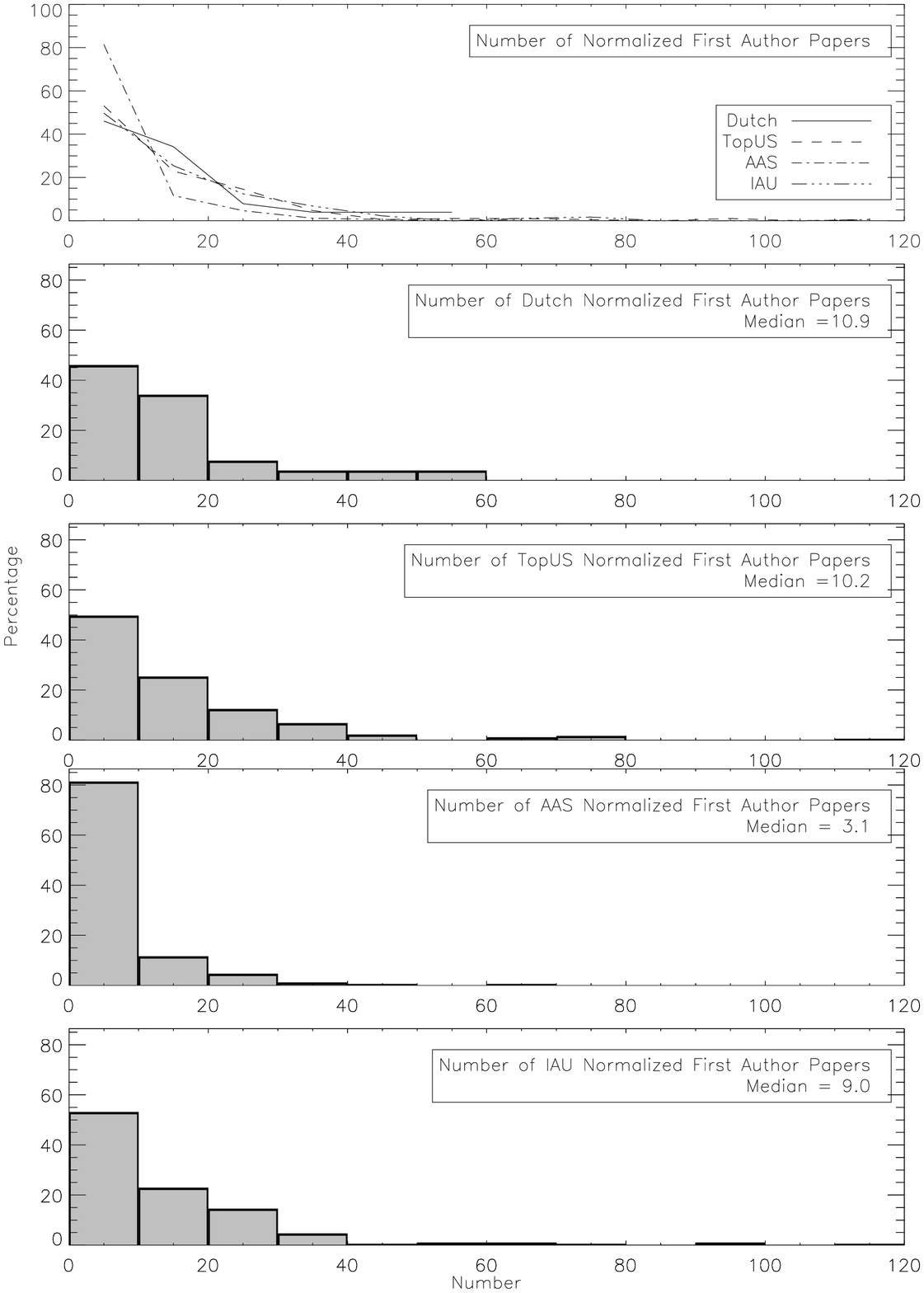}
\caption{\small As Figure \ref{fpapf} but now the papers are normalised to the number of authors.}
\label{fpapnormf}
\end{figure}

\begin{figure}[htbp]
\centering
\includegraphics[width=14.5cm]{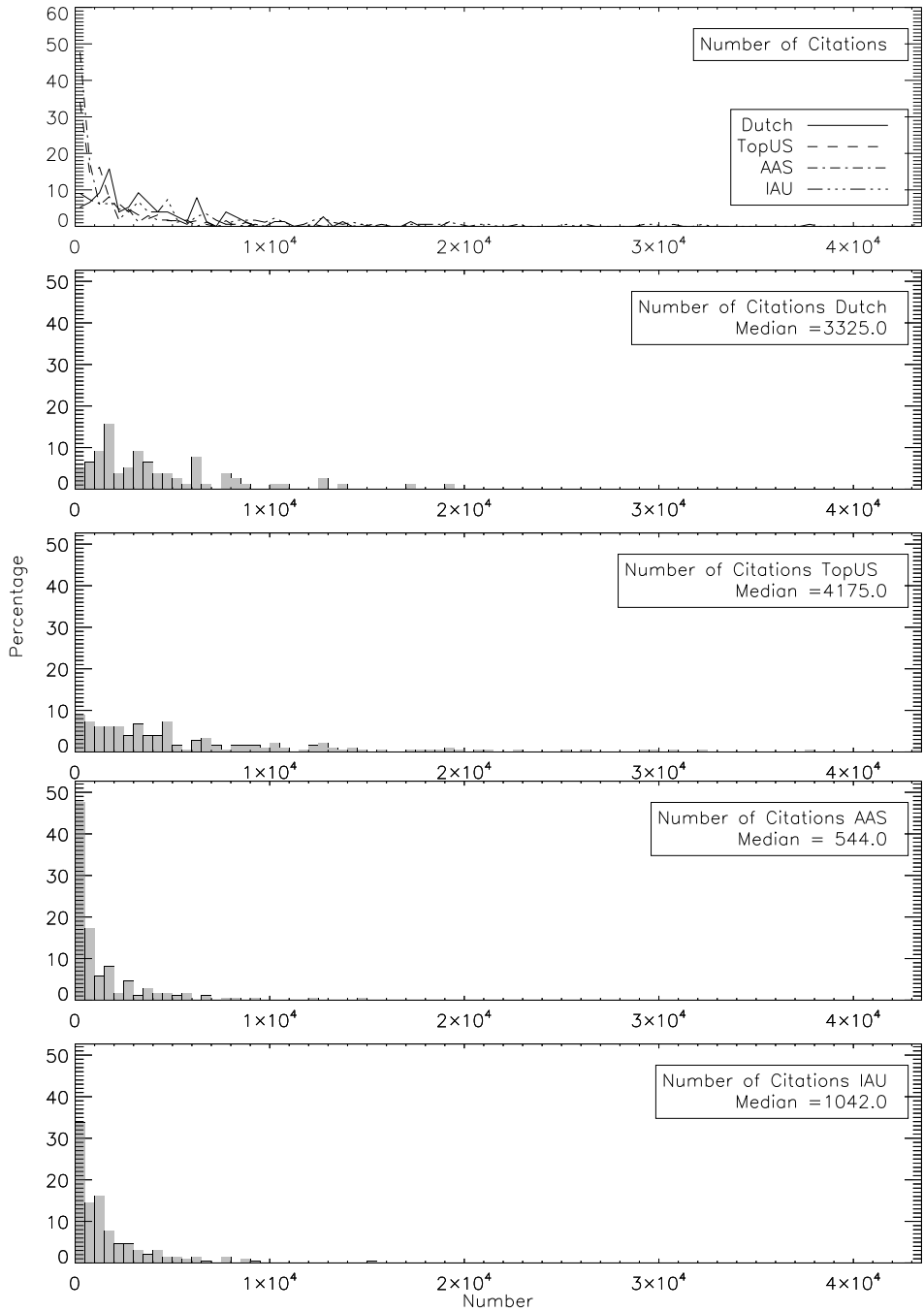}
\caption{\small As Figure \ref{fpap} but now for the citations received.}
\label{fcit}
\end{figure}

\begin{figure}[htbp]
\centering
\includegraphics[width=14.5cm]{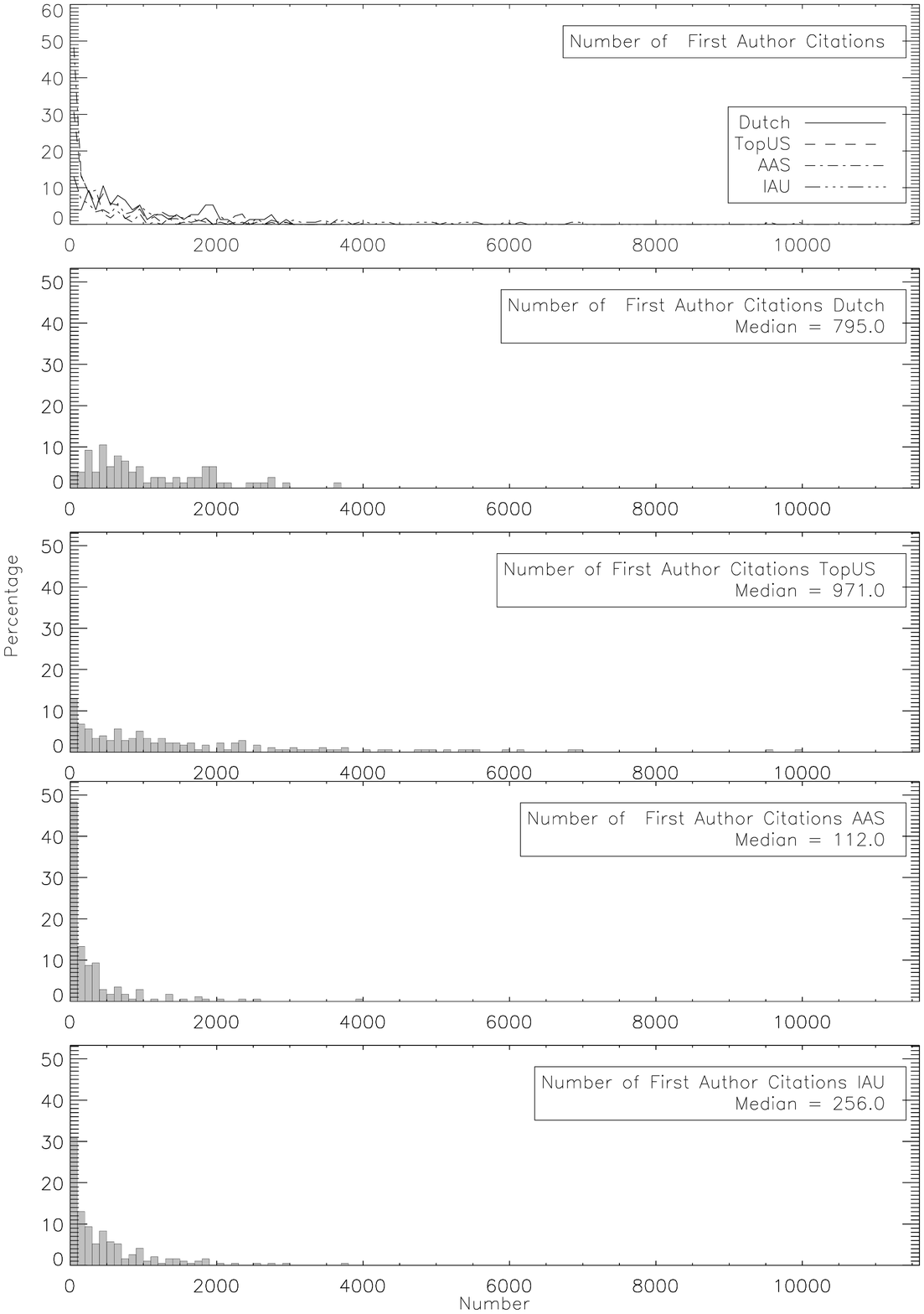}
\caption{\small As Figure \ref{fcit} but now only first author papers are considered.}
\label{fcitf}
\end{figure}

\begin{figure}[htbp]
\centering
\includegraphics[width=14.5cm]{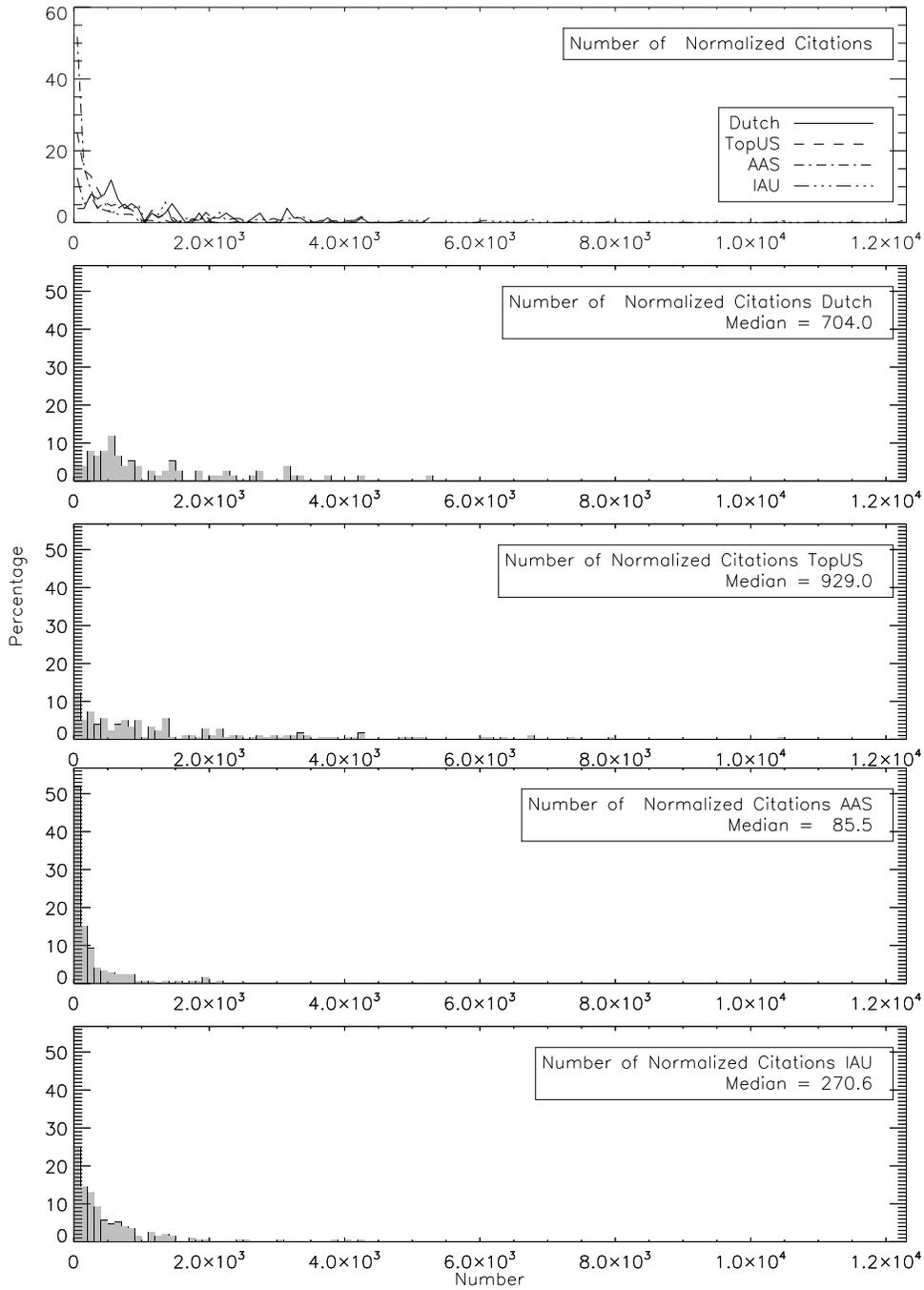}
\caption{\small As Figure \ref{fcit} but now the citations are normalised to the number of authors.}
\label{fcitnorm}
\end{figure}

\begin{figure}[htbp]
\centering
\includegraphics[width=14.5cm]{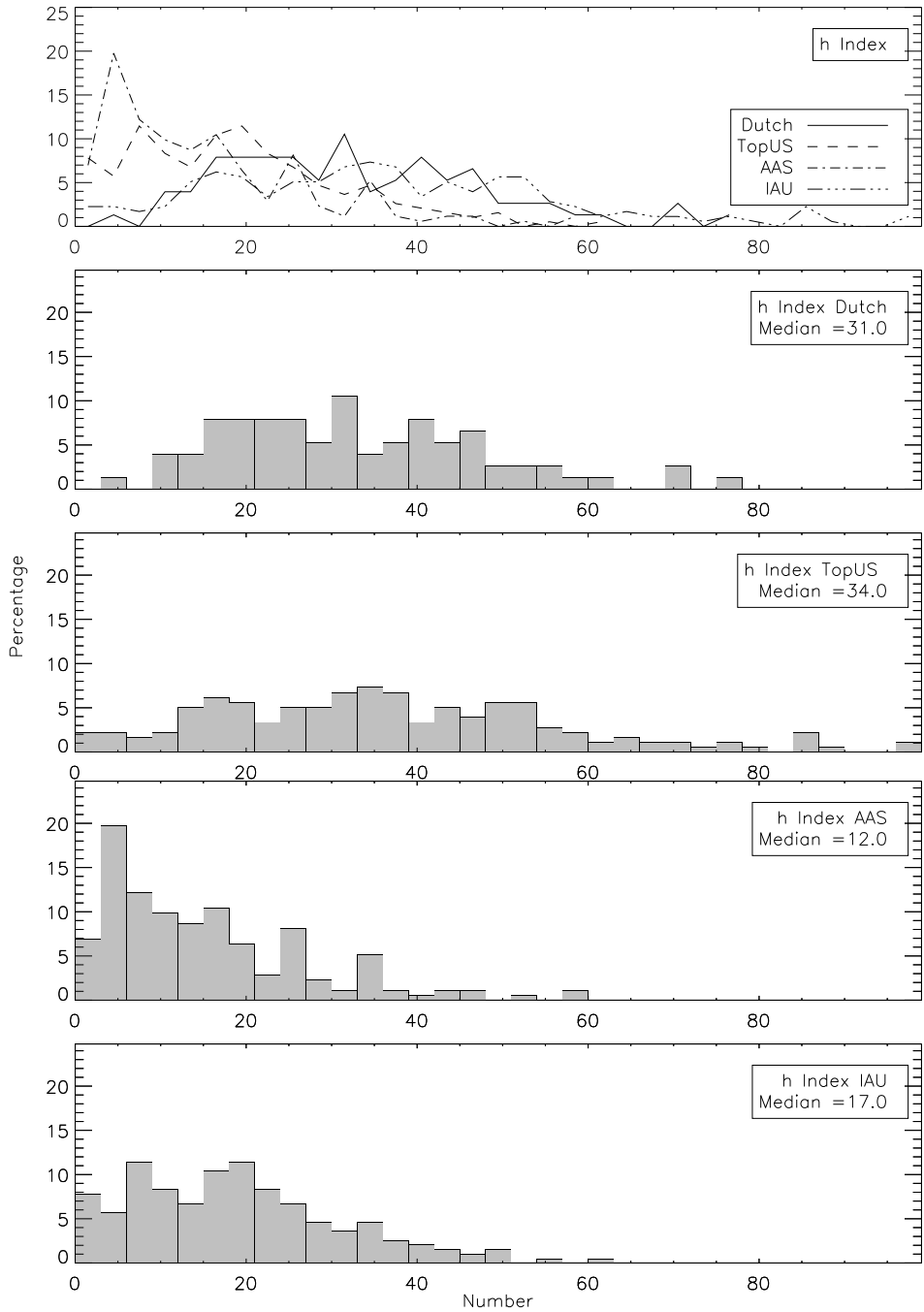}
\caption{\small As Figure \ref{fpap} but now for the $h$-index.}
\label{fh}
\end{figure}

\begin{figure}[htbp]
\centering
\includegraphics[width=14.5cm]{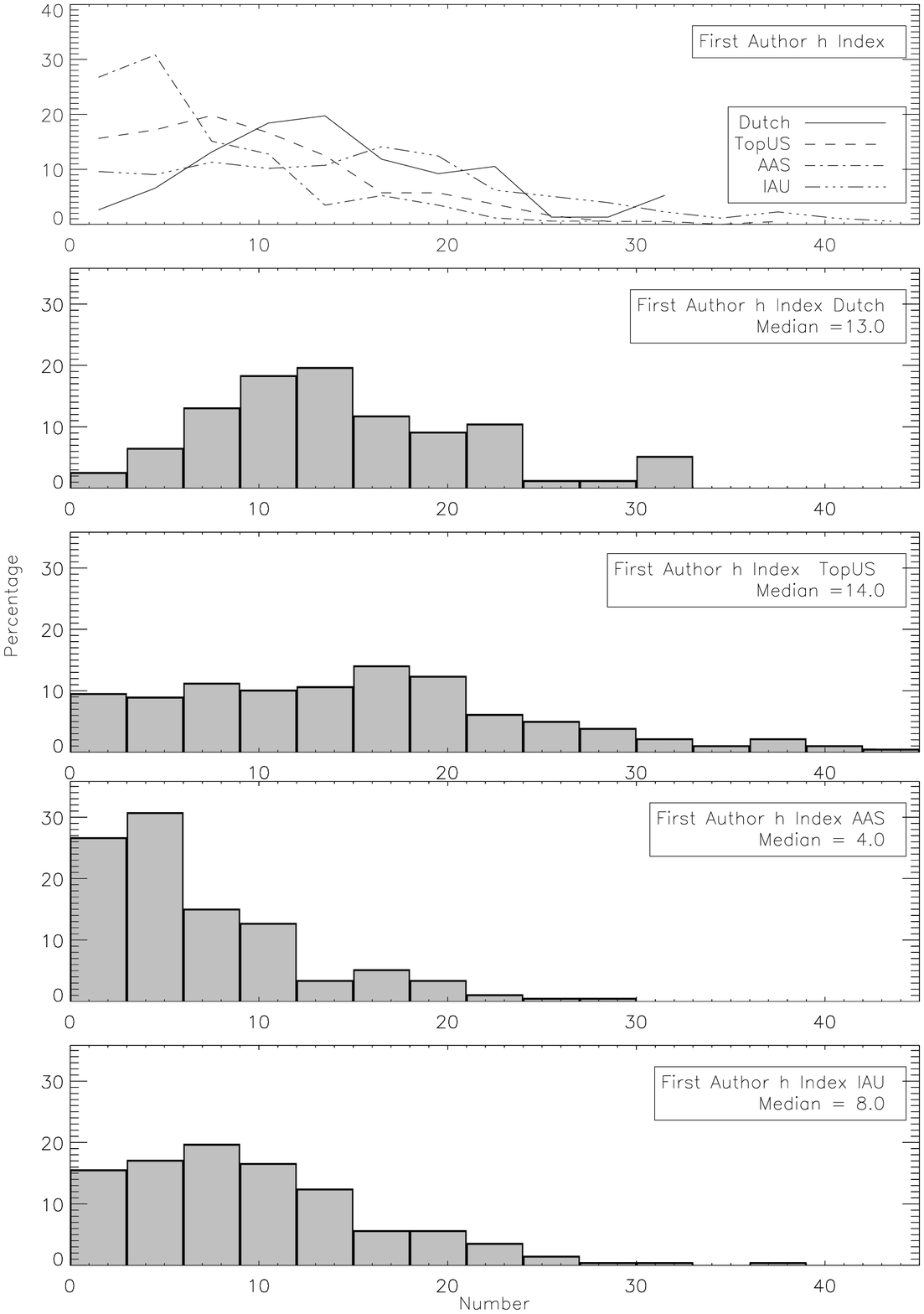}
\caption{\small As Figure \ref{fh} but now only first author papers are considered.}
\label{fhf}
\end{figure}

\begin{figure}[htbp]
\centering
\includegraphics[width=14.5cm]{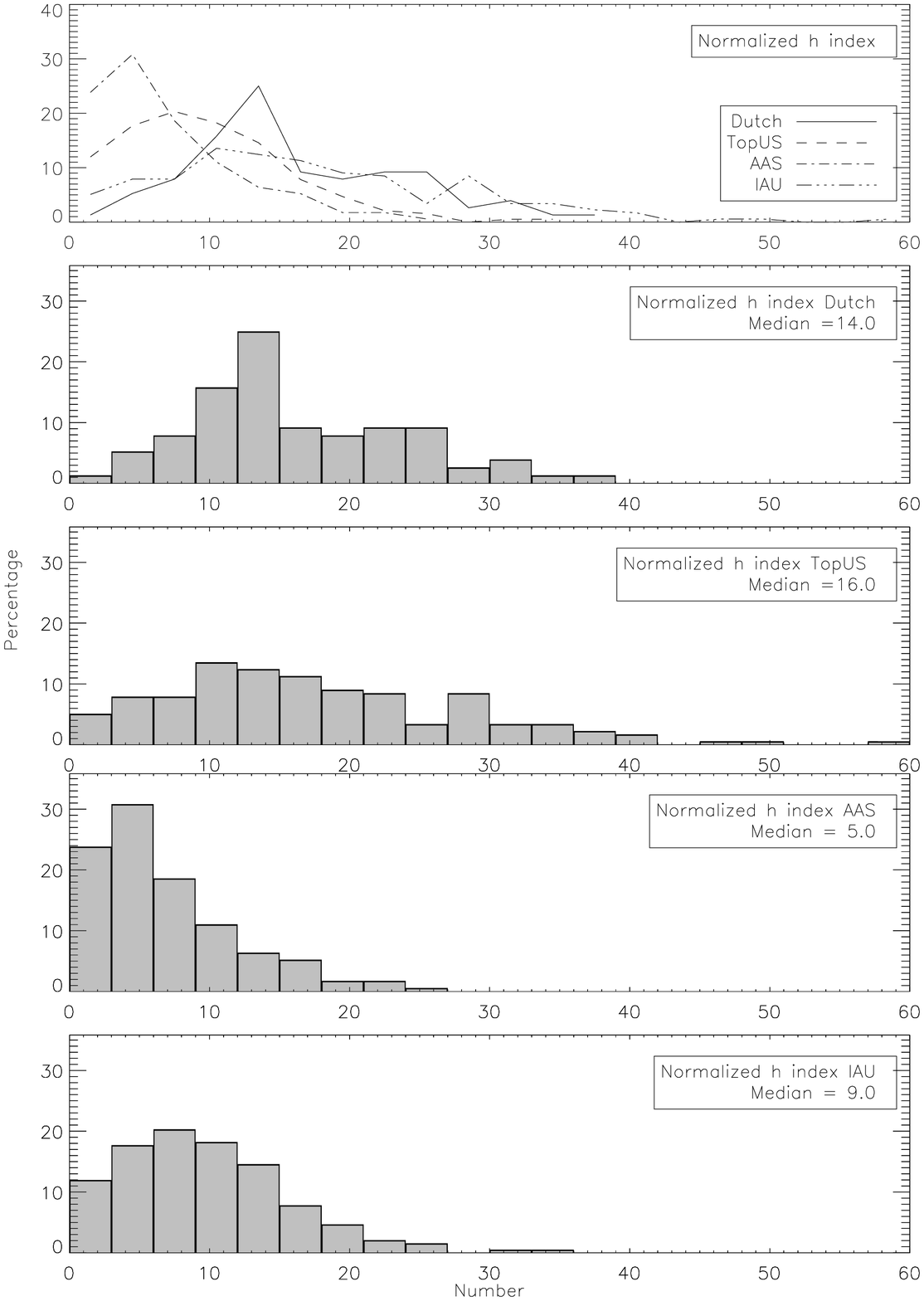}
\caption{\small As Figure \ref{fh} but based on normalised citations.}
\label{fhnorm}
\end{figure}

\begin{figure}[htbp]
\centering
\includegraphics[width=14.5cm]{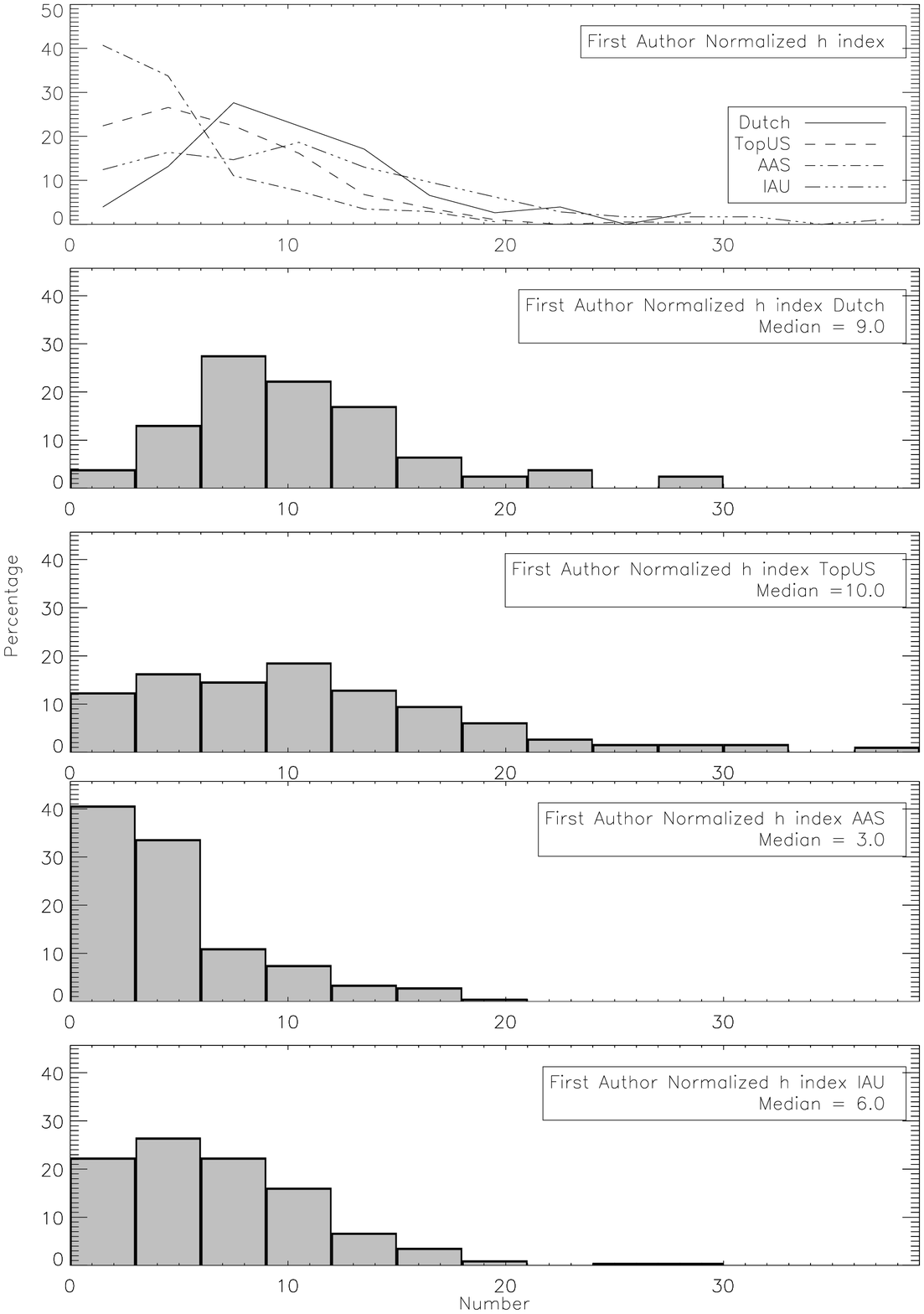}
\caption{\small As Figure \ref{fhf} but based on normalised citations.}
\label{fhnormf}
\end{figure}

\begin{figure}[htbp]
\centering
\includegraphics[width=14.5cm]{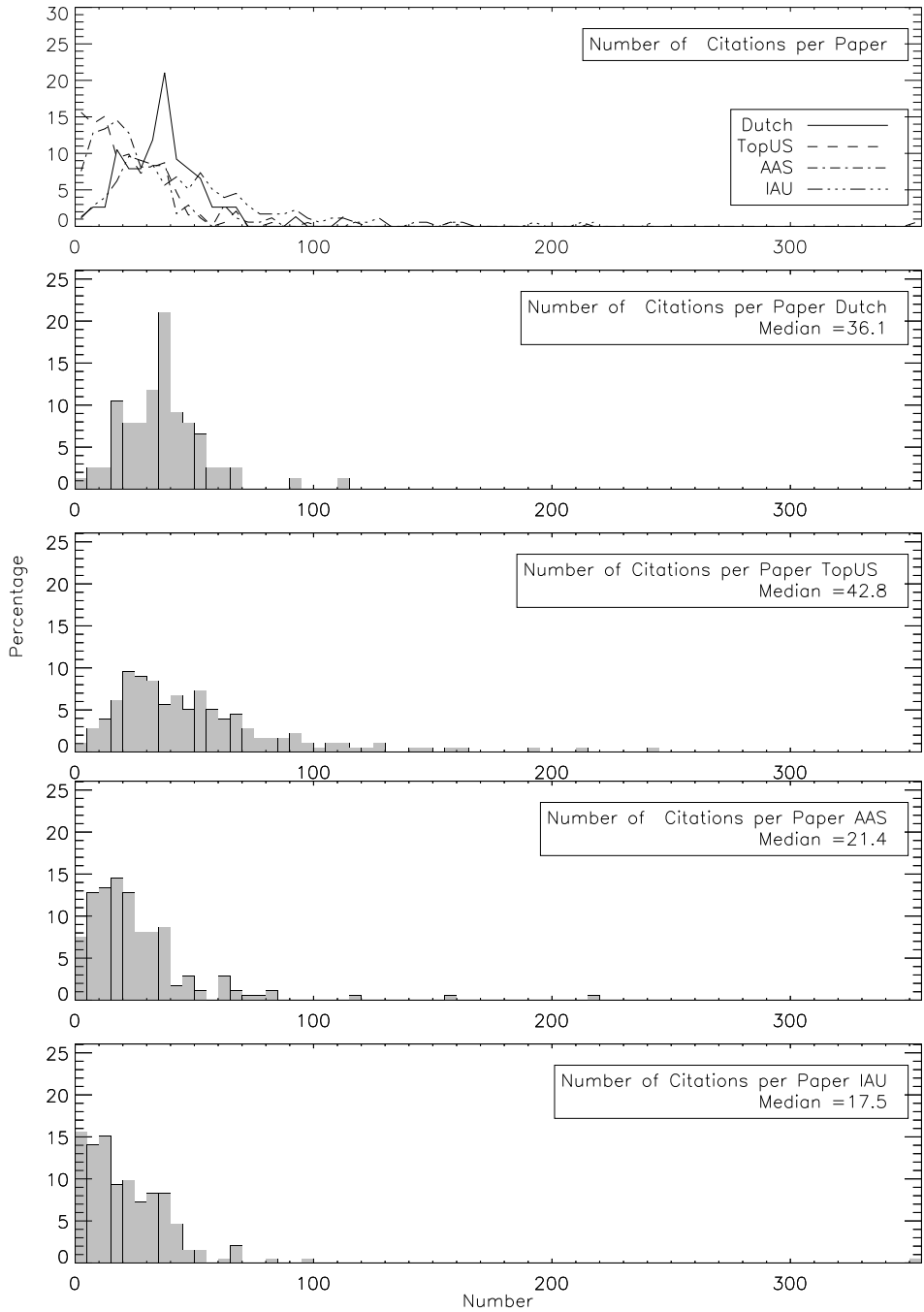}
\caption{\small As Figure \ref{fpap} but based on citations per paper.}
\label{fcitpap}
\end{figure}

\begin{figure}[htbp]
\centering
\includegraphics[width=14.5cm]{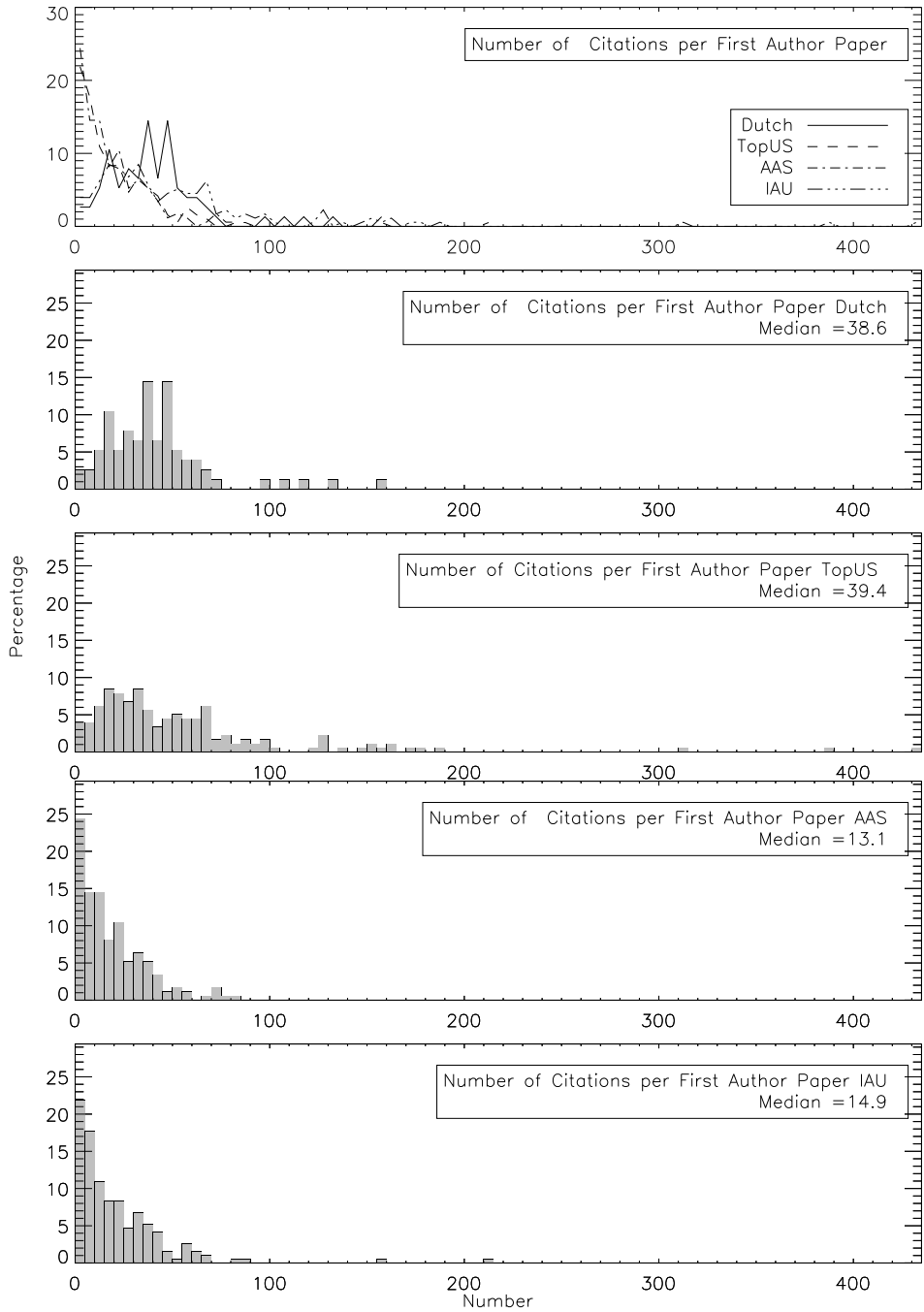}
\caption{\small As Figure \ref{fcitpap} but only considering first author papers.}
\label{fcitpapf}
\end{figure}

\begin{figure}[htbp]
\centering
\includegraphics[width=14.5cm]{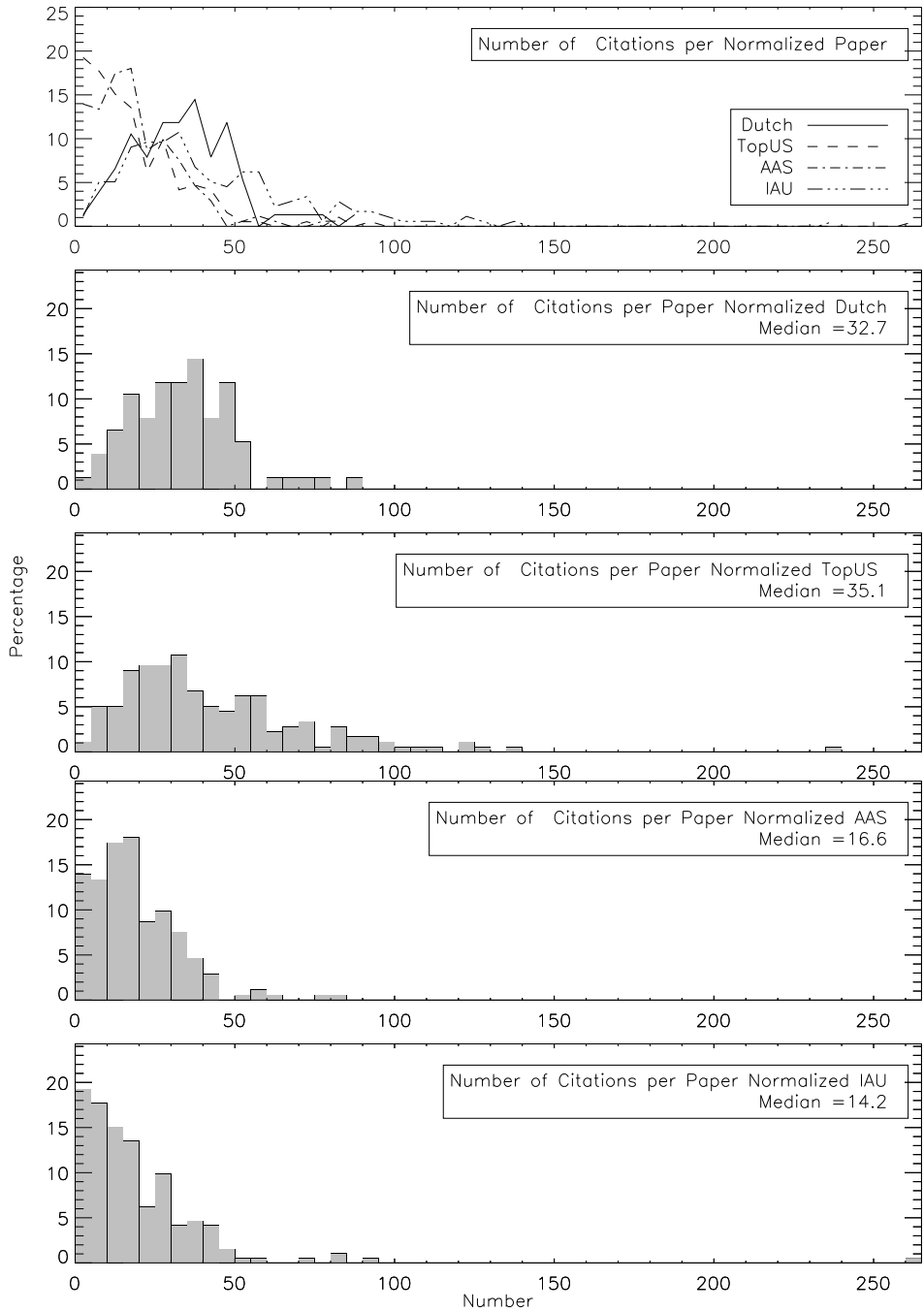}
\caption{\small As Figure \ref{fcitpap} but with citations and papers normalised to the number of contributing authors.}
\label{fcitpapnorm}
\end{figure}

\begin{figure}[htbp]
\centering
\includegraphics[width=14.5cm]{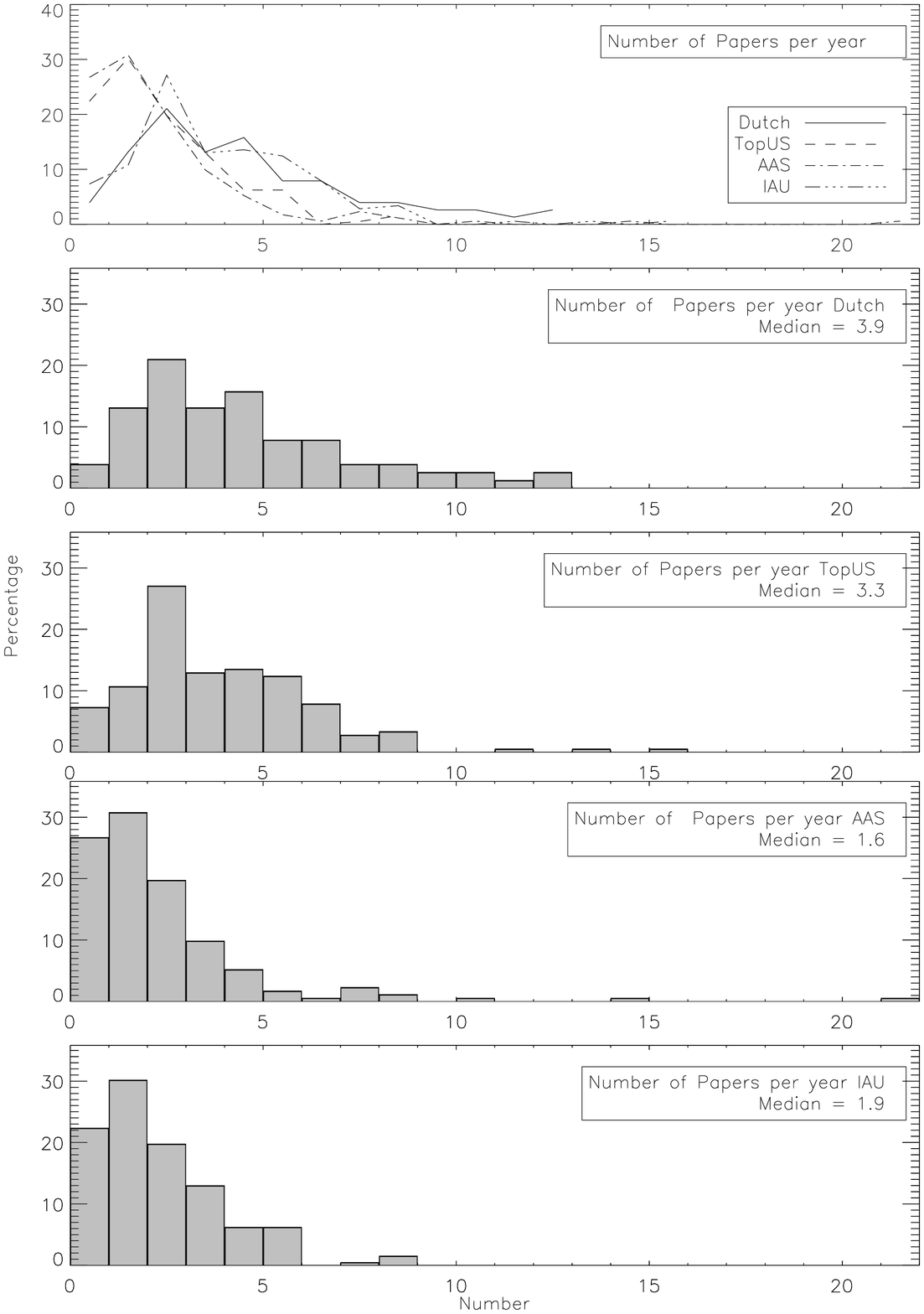}
\caption{\small As Figure \ref{fpap} but based on articles per year.}
\label{fyear}
\end{figure}

\begin{figure}[htbp]
\centering
\includegraphics[width=14.5cm]{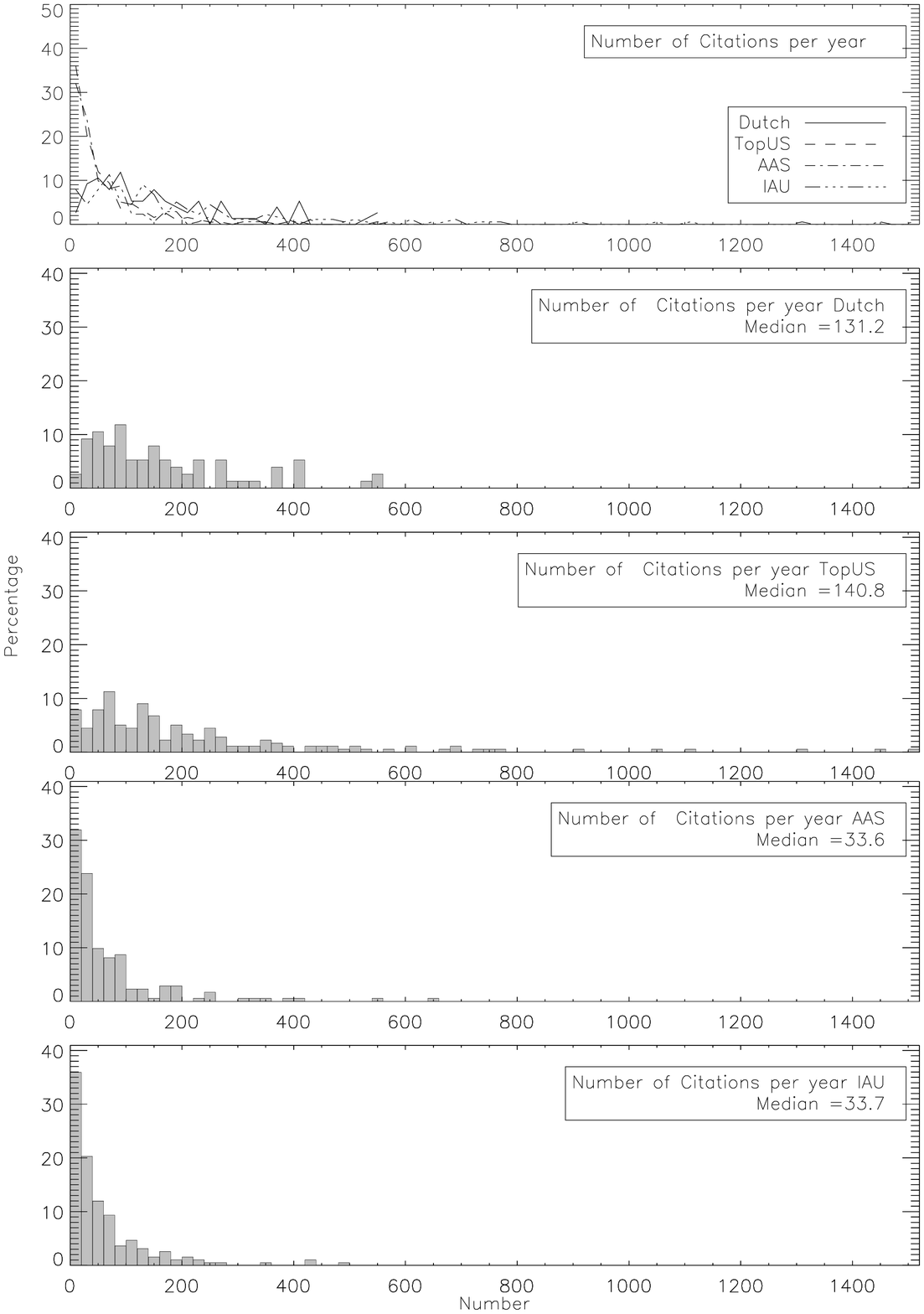}
\caption{\small As Figure \ref{fpap} but based on citations per year.}
\label{fyearcit}
\end{figure}

\begin{figure}[htbp]
\centering
\includegraphics[width=14.5cm]{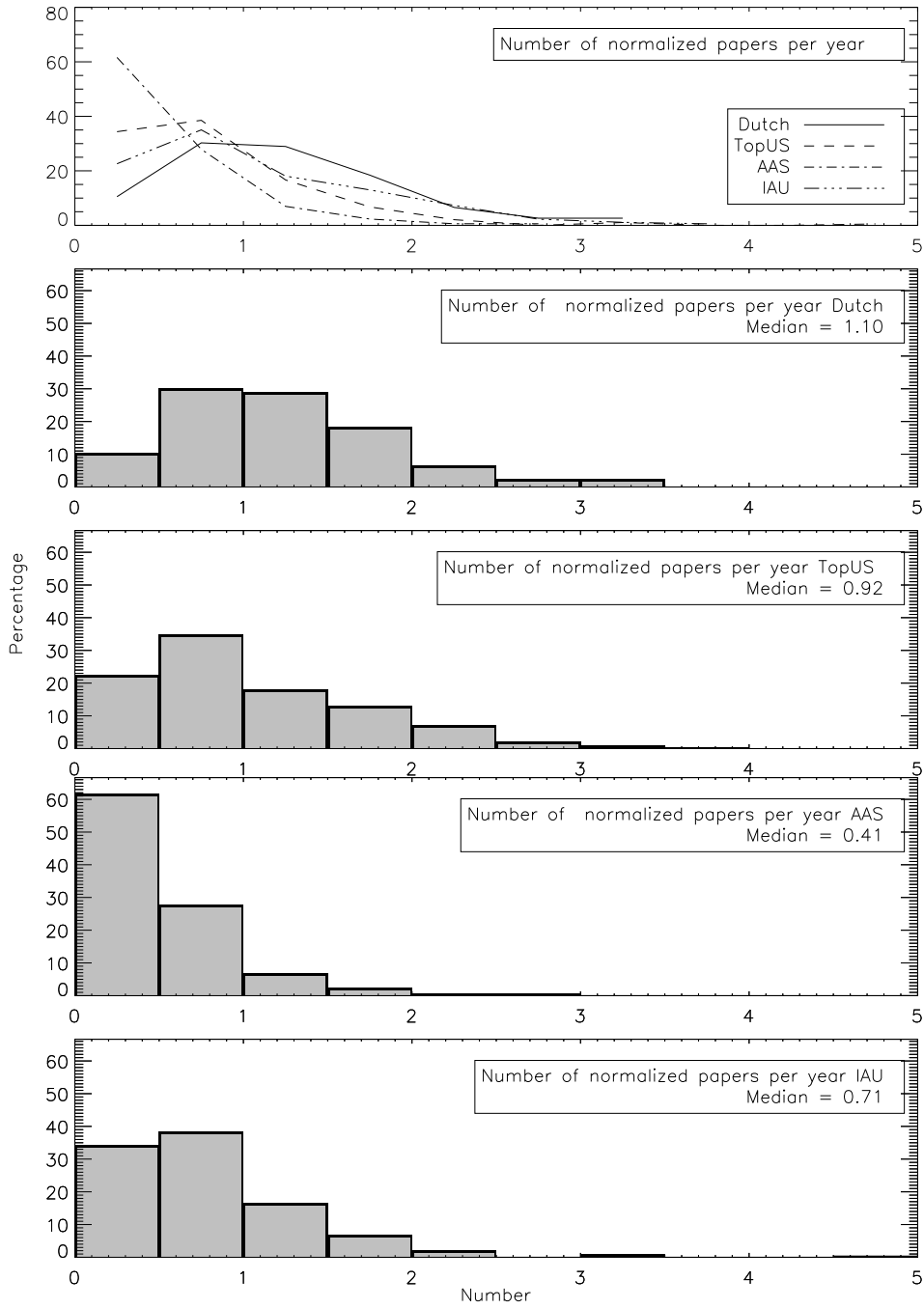}
\caption{\small As Figure \ref{fpap} but based on normalised articles per year.}
\label{fyearpapnorm}
\end{figure}

\begin{figure}[htbp]
\centering
\includegraphics[width=14.5cm]{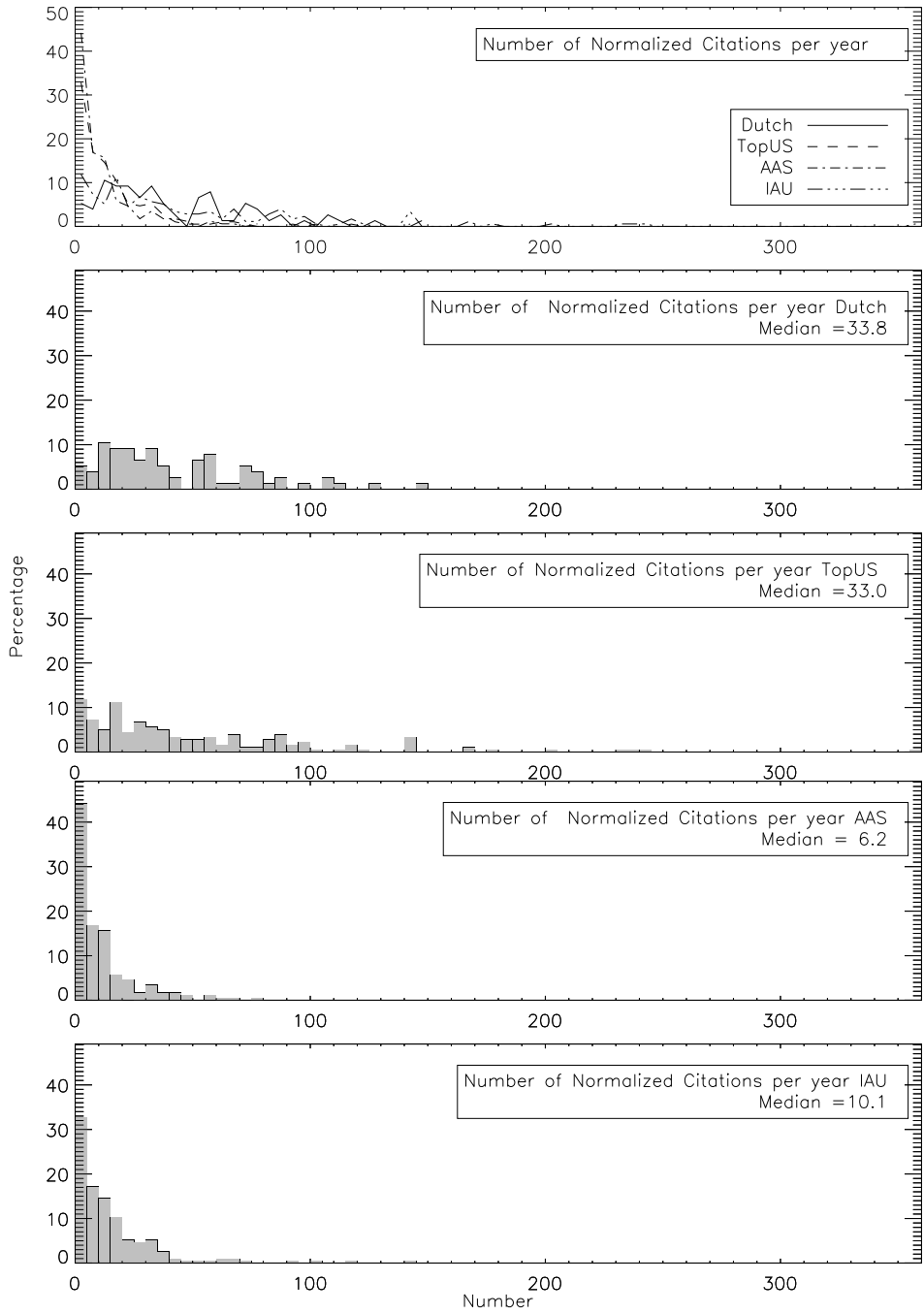}
\caption{\small As Figure \ref{fpap} but based on normalised citations per year.}
\label{fyearnorm}
\end{figure}

\begin{figure}[htbp]
\centering
\includegraphics[width=14.5cm]{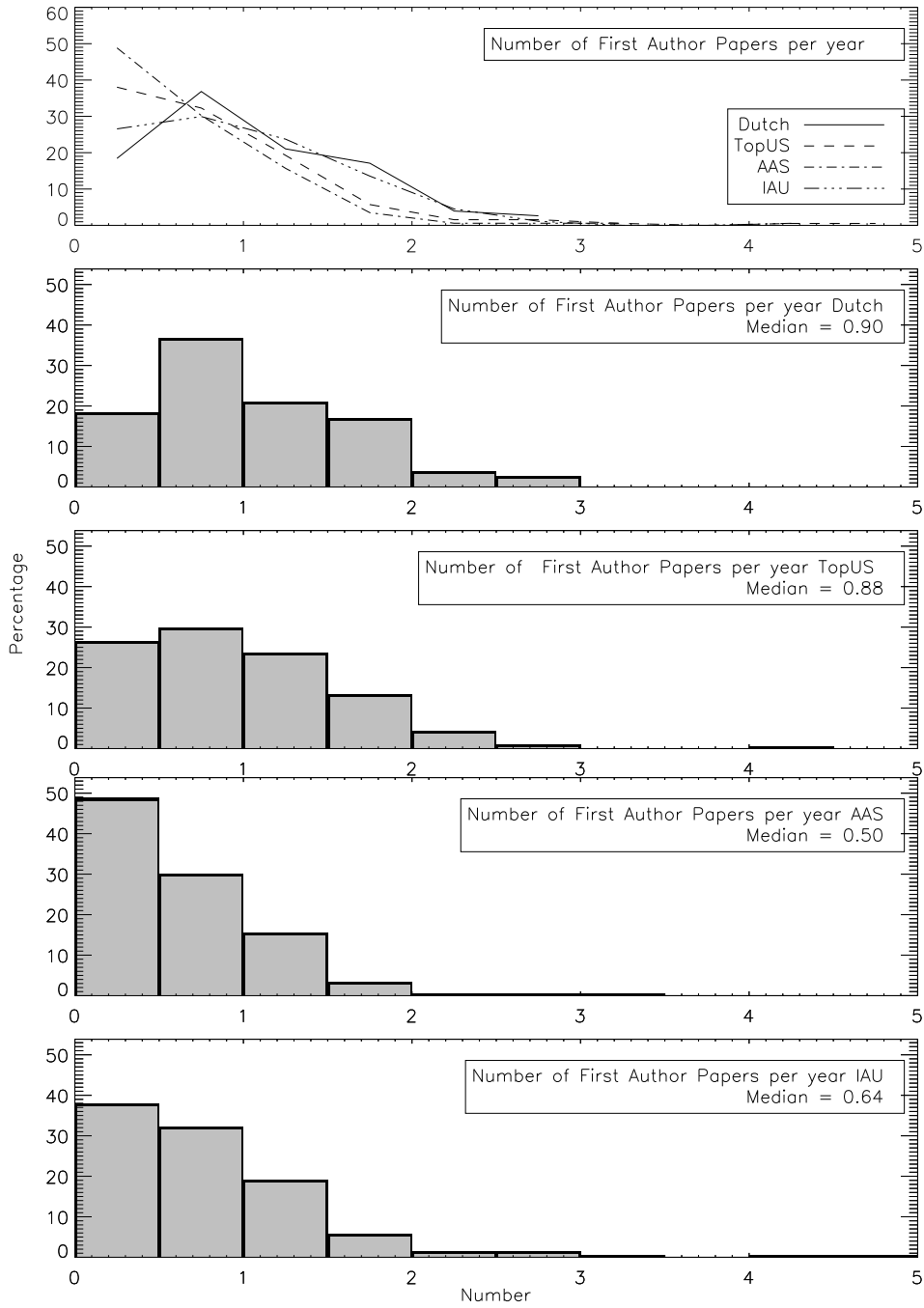}
\caption{\small As Figure \ref{fyear} but based on first author articles only.}
\label{fyearf}
\end{figure}

\begin{figure}[htbp]
\centering
\includegraphics[width=14.5cm]{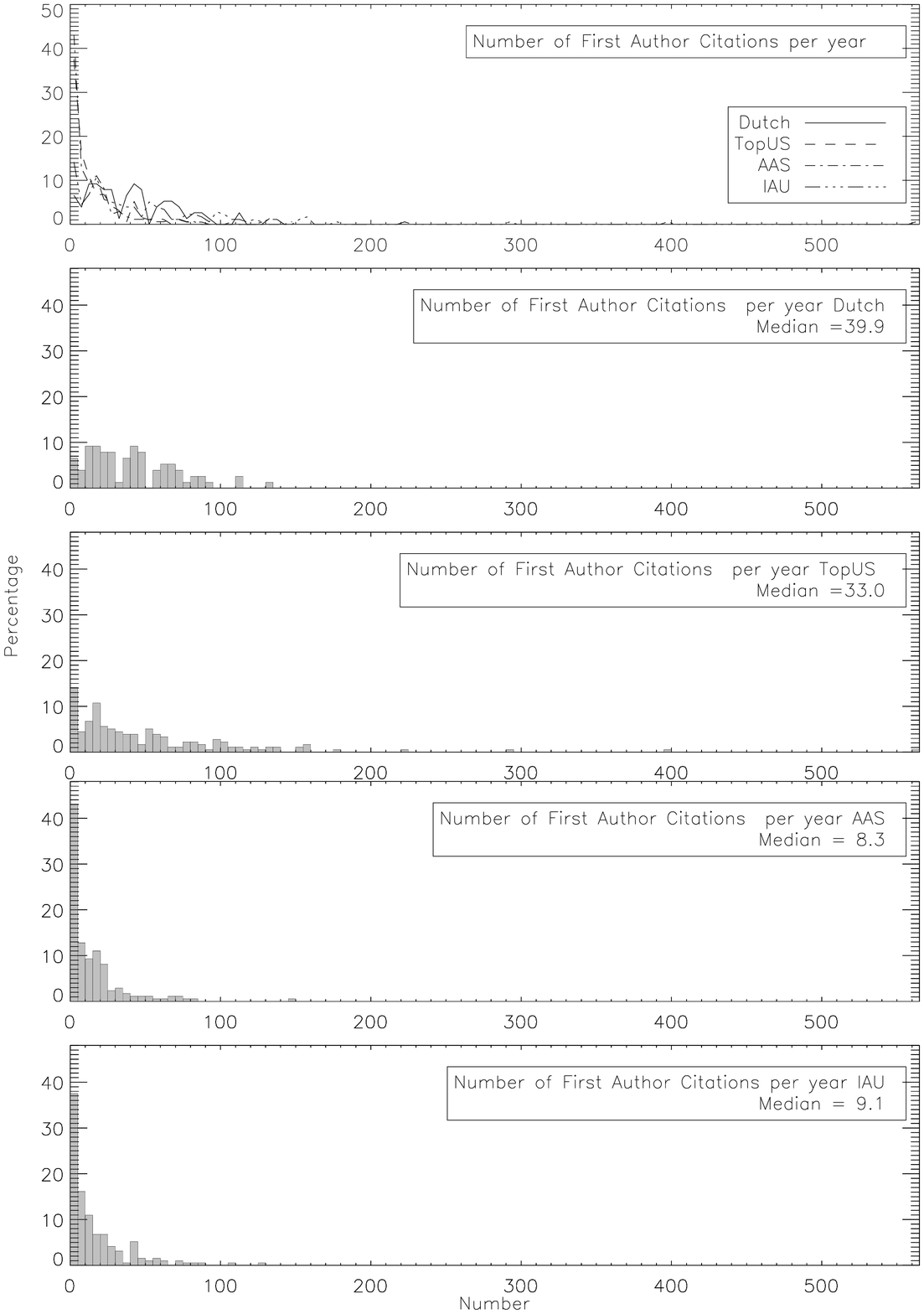}
\caption{\small As Figure \ref{fyearf} but based on citations per year.}
\label{fyearcitf}
\end{figure}

\begin{figure}[htbp]
\centering

\includegraphics[width=14.5cm]{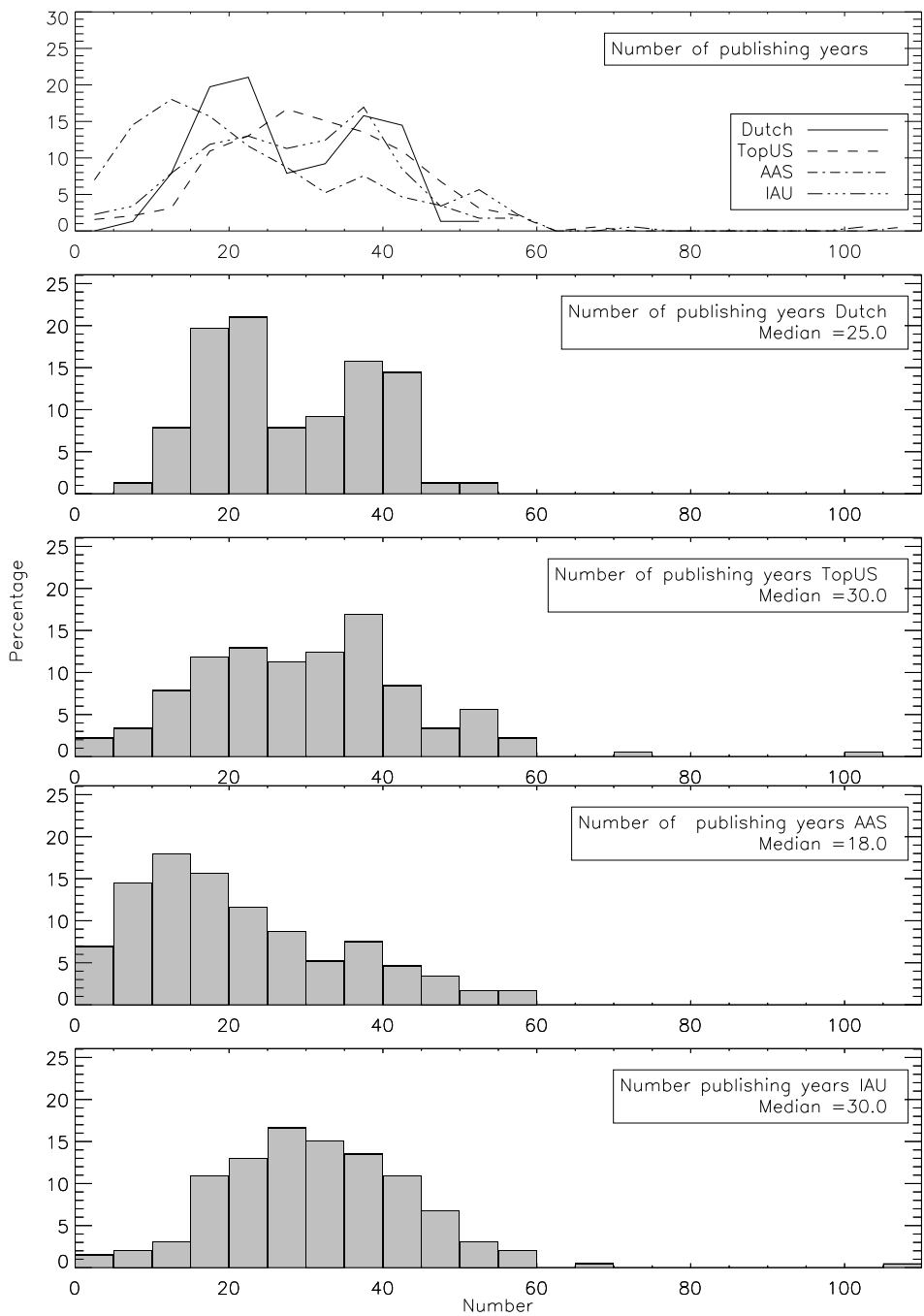}
\caption{\small As Figure \ref{fpap} but based on the number of years an author has been publishing (e.g. number of years between the first and last article).}
\label{fyearpubl}
\end{figure}

Figures \ref{fpap}--\ref{fyearpubl} show representations of  the
performance of  astronomers. Every figure shows 
in the bottom panels the histogram of astronomers of the TopUS, AAS and 
IAU samples. The second panel from
the top shows the same property for the Dutch (NOVA) 
staff members and the top panel compares the distributions of all samples
(NOVA -- solid line, TopUS -- dashed
line,  AAS -- dashed-dotted line. IAU -- dashed-three dotted
line). We collect the median values of the various distributions in 
Table~\ref{tabmed}.

Before discussion the results we want to stress that in comparing
the various samples it should be kept in mind that the selections are 
different for different samples. The NOVA and TopUS samples are active, tenured
staff members and are selected in a comparable manner. But the AAS and
IAU samples have been chosen at random from membership lists and will
contain in addition postdocs and other non-tenured astronomers, 
retired astronomers and persons that are associated with
astronomical research but to a large extent perform technical or other support 
functions. So, although the NOVA and TopUS samples are suitable for a direct
comparison, the AAS and IAU samples can be expected on the grounds of their
composition to score lower in bibliometric studies. They are useful, however, in 
finding out average (typical) values for astronomers of parameters such as the
$h$-index.

It is easily seen from Figures \ref{fpap} and \ref{fpapf} that the number of
papers produced by NOVA astronomers is clearly higher than that of astronomers
in the random samples and comparable to the astronomers in the top US 
institutions. However, it could be that that NOVA and TopUS astronomers have
on average a different number of  co-authors to their papers. 
This can be estimated from the ratio between the total and
normalised number of papers. For NOVA, TopUS and AAS this is between 3.5 and 4,
but for the IAU it is more like 2.9. In Figures \ref{fpapnorm} and 
\ref{fpapnormf} we see that the normalised number of papers for the 
IAU sample is more similar to NOVA and TopUS, but the AAS sample is not. 
IAU astronomers publish fewer papers, apparently with fewer 
co-authors than in the US and the Netherlands. 

The AAS sample gives surprising results. This sample performs at a  similar
or lower level than that for IAU membership. The cause must be that
there apparently are among the `active' AAS members many astronomers that
have not published very many papers. In fig.~\ref{fpapf} we see that for
the AAS astronomers about 45\%\ have fewer than 20 refereed papers. This
may seem surprising. However, a random check by hand, selecting only
those members whose names start with `Blo..', `Men..' en `Scha..'
resulted in 14 out of 36 (39\%) having 20 refereed papers or less and
9 (25\%) more having between 21 and 40 papers. The number of papers per
year is similar to that of the IAU sample. It is an effect of age (see 
Table 1):
AAS astronomers have published on average for fewer years than the 
other samples. There
appears no such difference between the NOVA and AAS samples; the 
NOVA astronomer 
on average  publishes more papers per year. In fact, the 
average number of
years that an astronomer has been publishing is 29 for NOVA and 
TopUS and 30 for IAU,
but only 18 for the AAS sample (Figure \ref{fyearpubl}). 
It does not have much to do with people
publishing papers with many authors on them. The number of normalised
papers per year is 0.9 (NOVA), 0.9 (TopUS), 0.4 (AAS) and 0.7 (IAU), 
so there 
appears to be less production in the AAS sample per astronomer. This was not
found in a related older paper\footnote{{\it ``A comparison of astronomy
in fifteen member countries of the Organisation for Economic
Co-operation and Development''}, P.C. van der Kruit, Scientometrics 31, 155-172
(1994), and {\it ``  The astronomical community in the Netherlands''}
Quart. J. R.A.S. 35, 409 (1995). Available at 
www.astro.rug.nl/$\sim$vdkruit/\linebreak[4]
jea3/homepage/oecd.pdf and
www.rug.nl/$\sim$vdkruit/jea3/homepage/qjras.pdf.}.

>From Figures \ref{fcit}-\ref{fcitnorm} we see that the number of
citations and related parameters
for NOVA astronomers is much higher than that of the IAU and AAS
samples. 
One can also imagine that the distribution of citations per paper differs
between the samples. We therefore show the the citations per paper 
in Figures \ref{fcitpap}-\ref{fcitpapnorm} and the $h$-index 
(Figures \ref{fh}-\ref{fhnormf}). We also compare the 
citations received per year (Figures \ref{fyear}-\ref{fyearcitf}). 
All these figures show that the number of citations to papers  
by NOVA astronomers is significantly higher than that in the
samples randomly selected randomly from the IAU and AAS membership lists. 
We noted already that this comparison is not entirely fair, as the
NOVA-sample consists exclusively of tenured, active researchers.

The comparsion of NOVA researchers to the faculty of the top-15 institutes 
in the USA, however, is between two samples that are selected similarly
and the two samples are directly suited to look for differences in 
bibliometric performance. It is obvious from Table~\ref{tabmed} that the NOVA
sample performs as good or almost as good as the TopUS astronomers. Note
that the number of publishing years is less for the NOVA sample than for the TopUS
astronomers, indicating a smaller average age of Dutch tenured staff. Indeed
recently (to a large extent due to the NOVA funding from the Bonus-incentives 
Scheme) a large number of new hires have been made, often replacing retiring 
staff on so-called `overlap' positions. 
The set of key-researchers does even better, but these are of course 
selected from the most senior astronomers (none of them has an $h$-index below
20) and this sample should not be compared directly with the TopUS staff. 
\bigskip

{\bf Summarising:}\\
{\it We find that the NOVA researchers perform much better among
bibliometric measures than samples drawn from IAU or AAS membership
lists. More suitable for a comparison is with the (tenured) staff of the
top-15 US institutions and there the outcome is that NOVA staff performs in
these respects as good or almost as good as that of American top institutes.}

\section{The published NOWT results}
\hspace*{0.5 cm}Every three years the level of Dutch research and development 
is compared to that of other OESO countries. This is done in the
{\it Nederlands Observatorium van Wetenschap en Technologie} (NOWT),
the {\it Netherlands Observatory of Science and Technology}, 
which regularly publishes
a report {\it Science and Technology Indicators}. 
This study is commissioned by the
{\it Ministery of Education, Culture and Science} (OCW) 
and performed jointly by the
{\it Center for Science and Technology Studies} (CWTS) of Leiden University and 
the {\it Maastricht Economic and social Research and training centre on
Innovation and Technoly} (MERIT) of Maastricht University. 

The NOWT makes use of several indicators. One of the most important ones
is the number of citations to papers in refereed journals normalised by the number of 
citations that the average articles from the same years receive. 
This parameter is called the
{\it impact ratio} and it indicates whether an article receives more 
citations (impact ratio $>$ 1) or less (impact ratio $<$ 1) than 
the average paper worldwide in the same discipline and published in the same 
year. This is used as an important indication of the quality, visibility
and impact of scientific disciplines and the years preceding the report.
Usually the window of the years
of publication of the papers and the citations is 3 or 4 years prior
to that of the NOWT study. 

Between the reports of 2005 and 2008 the impact ratio of Dutch 
astronomy as a whole dropped from 1.25\footnote{Or 1.27; NOWT  gives 
different results in their tables 4.6 and 4.7 of the 2005 report} to 1.19
according to NOWT. 
In the following we will investigate the reasons of this drop and whether this
conclusion is robust. 
This will be done by analysing the citation patterns of
articles with authors from the institutes that make up the {\it Netherlands 
Research School for Astronomy} (NOVA), using as a tool the {\it NASA 
Astrophysics Data System} (ADS).
 
We will examine first the results published by NOWT. 
We note first that between the reports there is an important
difference in the focus of the studies; the 2005 report paid special attention
to the workforce in R\&D, whereas the focus of 2008 was 
budgets. Within the area of bibliometric parameters there is also a difference
in emphasis between subsequent NOWT reports.

When we look at the impact ratios, we see that in the 2008 
NOWT this is presented as the {\it total impact ratio} of 
the universities\footnote{This is the impact ratio of 1.19 mentioned as the 
result for astronomy 2008.} as well as separate scores for the five 
universities and of the institutes of ASTRON and SRON.\footnote{The 
{\it Netherlands Foundation for Research in Astronomy} ASTRON and the 
{\it Foundation for
Space Research in the Netherlands} SRON are funded by the {\it Netherlands 
Organisation for Scientific Research} (NWO) primarily as technological 
organisations; they have institutes in respectively Dwingeloo and 
Utrecht/Groningen primarily for the  design and construction of scientific 
instruments and the operation of observing facilities. 
Although  there definitely is also related astronomical research, the
staff of these institutes are not part of NOVA, which is a federation
of astronomical institutes at \underline{universities}. 
On the other hand, most astronomers employed by ASTRON and SRON have additional
parttime or unpaid affiliations with NOVA institutes.}
The contributions of  ASTRON and SRON are 
not all purely astronomical, but it is not a priori clear how NOWT treats 
this. Another thing that is striking in 
the 2008 report is that NIKHEF (NWO/FOM institute for high energy physics), 
Rijnhuizen (NWO/FOM institute for plasma physics) and KNMI (meteorology) 
are considered to produce astronomical publications. This may very well
be unjustified; in any case we usually do not regard  
the staff of these institutes as part of the Dutch astronomical community. 
Certainly they are not part of the NOVA federation of institutes.
Where the Rijnhuizen and KNMI 
effects are small and probably negligible (19 and 10 papers respectively), 
the NIKHEF contribution is considerable (118 articles or of order 10\%).

In the 2005 report the only number 
reported is an impact ratio for Dutch astronomy as a whole, being either 
1.25 (Table 4.6) or 1.27 (Table 4.7). This raises some uncertainties. It is
for example not clear whether or not this includes publications from
the  ASTRON and SRON institutes\footnote{It is true that separate results 
are given for ASTRON as an institute.}. 
And the question is if it also includes NIKHEF,  Rijnhuizen and 
the KNMI? 

Going back further in time to the 2003 report, we see that in 
that report the total impact ratio of Dutch astronomy is 1.29.  This 2003 NOWT
also provides the impact ratios of the different institutions separately. 
In this report the impact ratio of astronomy at the universities is equal to 
the total impact ratio (1.29). It is interesting to see that in 2003  the 
contributions from ASTRON and SRON are not split into different disciplines 
and therefore one would assume that they are  considered to be completely 
astronomical.

Table \ref{tabnum} shows the results retrieved from the NOWT 
reports back to 1998. From this table we see that 
the impact ratio in 1998 is lower than the current one; however, only 
two years later the highest impact ratio of the last decade is reported.
So there are fluctuations on very short timescales that are unlikely to 
reflect significant variations in productivity, quality or relevance of 
Dutch astronomal research. 
Also, in 2000 and 2008 a total number for the universities only (essentially
the NOVA federation) is given, 
e.g. not including ASTRON, SRON and other possible non-academic
institutes.

\begin{table}[t]
\centering
\begin{tabular}{ccc}
\hline
year& impact & Univ.\\
\hline
2008&1.19&Y\\
2005&1.25&\\
2003&1.29&\\
2000&1.30&Y\\
1998&1.07&\\
\hline
\end{tabular}\caption{\small  Impact ratios for Dutch astronomy from NOWT reports. 
The last column indicates whether or not the number applies solely for the 
institutes at the universities}
\label{tabnum}
\end{table}%

The astronomy group at the Radboud Universiteit Nijmegen (RU) 
is new and included in the NOWT report for the first time in 2008. If we 
look at the impact ratio cited in the report of 2008 we see that their impact 
according to the NOWT is low (impact ratio = 1.03) compared to the other 
universities. They produce a much lower quantity of articles which reduces 
their contribution to the total number. In fact the impact ratio for the 
universities excluding RU would be 1.20.\\
\hspace*{0.5 cm}{\it These results produced in the NOWT reports 
suggest that the difference in impact ratio between 2005 and 2008 is within 
the normal scatter caused by natural fluctuations as well as inconsist 
methods of calculating the 
actual number for the whole of Dutch astronomy; the same must hold  for
the impact ratios of individual institutes.}

\section{The NOWT studies repeated using ADS}
\hspace*{0.5 cm}ADS has a query form where articles can be selected according to
author, set of authors, words in title or abstract, journal, year of publication,
etc. It also has an unsupported query form to search by 
affiliation, which we will use below.  
Since this is an unsupported feature of the database services,  it is important
to establish the trustworthiness of this type of query.
To investigate the reliability of ADS we have obtained all the articles of
the Kapteyn Astronomical Institute as well as those of the Sterrewacht 
Leiden and compared them to the Annual reports of these institutes for 
the years 1998, 2000, 2003, 2004, 2005 and 2006. From this we find that 
ADS can be used as a reliable source back to the year 2000. Before 2000 
entries for the major  journals {\it Astrophysical Journal} and {\it Astronomical 
Journal} do not list  affiliation consistently in ADS.

>From the comparison of the Kapteyn Institute it became 
clear that persons with a cross-affiliation are often identified with 
their main affiliation only. Therefore in ADS roughly 15\%\ of articles 
listed in the Annual reports are not attributed to the Kapteyn Institute
since their authors have an unpaid-appointment with Groningen University
in addition to their main affiliation (often ASTRON or SRON)
and are solely affiliated  with their main institutions in ADS. 
This problem should also affect the ratios presented in the NOWT
reports, but is irrelevant for the impact ratio of  Dutch astronomy as a
whole.  Another mismatch between the lists retrieved from ADS and the 
Annual Reports is caused by authors who recently have been relocated to 
a different institute. These authors are often affiliated with their new
institutes while much of the work for the article concerned  was done at
their previous location. Fortunately, this behaviour seems to produce 
a similar number
of additional articles as missing ones when compared to the annual reports.
Of course, the actual affiliation mentioned on the papers corresponds to that 
found in the ADS listing and not to the Annual Reports.  Even though there are 
cases where the affiliation mentioned on the paper is incorrect, these cases 
seem to be rare and they would most likely not be corrected  by CWTS.

We therefore conclude that back until the year 2000, 
the affiliation search of ADS can be used to construct correct publication 
lists of the Dutch institutes. This period (2000-2008) also covers the period 
of the so-called {\it Dieptestrategie}\footnote{The {\it `Bonus-incentives 
Scheme'},
in which NOVA has been recognized as a top-research school in 1998 and 
receives additional funding since then.}
\bigskip

We have obtained from ADS lists of all refereed articles, which 
are affiliated  to one of the Dutch institutes, plus the number of citations 
to that article received up to December 2008.  The number of citations of
a paper here is the number of references to that article in {\it all} 
bibliographic publications, refereed as well as the unrefereed. 
We have calculated the impact ratios of the institutes by normalising
these counts by average citation rates of papers in the same years of
publication and with the same citation window. We have done that 
in several ways. 

First, we determined impact ratios by taking only those
Dutch publications that appeared in one of the four major astronomical journals
and normalised these with the average number of citations of all papers
in these journals published in the same years. 
The journals are {\it Astrophysical Journal, Astronomical Journal, 
Monthly Notices of the Royal Astronomical Society} and {\it Astronomy and 
Astrophysics}.  Secondly, we did the same excercise, but now included in addition
the astronomical publications in {\it Science} and {\it Nature}. Even though the 
number of articles in these latter journals is low, their impact is generally 
extremely high.  Thirdly, we include two smaller journals (namely 
{\it Astronomische Nachrichten} and {\it New Astronomy Reviews}) in order
to see whether the addition of such journals severely affects the impact 
ratios. Fourthly, we calculated the citation rates for the total of
all refereed Journals in ADS 
as listed in {\tt http://adsabs.harvard.edu/abs\_doc/refereed.html}. 
In this case of `all journals' the numbers involved are so large that ADS 
can only handle one month at the time. Therefore it is very
time-consuming to 
calculate the normalisation factor. We have solved this by taking 
only the first and last two months into consideration. This assumes that 
the number of citations per article declines linearly throughout the
period considered.

\begin{table}[!tb]
\begin{center}
\begin{tabular}{lllll}
\hline
Method& 2000-2002 & 2000-2003 & 2003-2005 & 2003-2006\\
\hline
4 Major journals&1.22(873)&1.21(1193)&1.27( 970)&1.27(1399)\\
6 Major journals&1.23(889) & 1.24(1219) & 1.31(1182) & 1.31(1428)\\
8 Journals&1.27(893)&1.27(1238)&1.31(1024)&1.31(1490)\\
All refereed journals&1.87(986)&1.89(1376)&2.00(1182)&2.00(1721)\\
\hline
Citation window&1.81(986)&1.78(1376)&1.93(1182)&1.87(1721)\\
All NL & 1.74(1285) & 1.71(1830) & 1.86(1578) & 1.80(2261) \\
NOWT & &1.27(1807)& & 1.19(1311/1407$^{*)}$)\\
\hline
\end{tabular}
\end{center}
\caption{\small  Impact ratios (Number of publications) for the 
different methods described in the text
for the periods 2000-2002, 2000-2003, 2003-2005, 
2003-2006 (always last year inclusive). In the top part, the impact ratios are 
the summed results of the Dutch universities together and the citation window
extends to (and including) 2008. In the bottom part the publication and
citation windows are the same. The first row is for the universities
together and the second one all Dutch publications. The NOWT impact ratios
in the bottom line are for all Dutch publications for 2000-2003 and for
the universities only for 2003-2006.}
{\small
*)The publication output is 
given as a percentage of the total output in table 4.6 of the 2008 NOWT 
report. The individual results of the universities amount to a higher 
total than the percentage of total output given in this table. }
\label{tabnumuni}
\end{table}

To make a comparison to the NOWT results, 
we finally used a citation window equal to that of  
the publication years, as is done in the CWTS and NOWT studies. We used again
all refereed journals. 
Such a citation window is uncomfortably small since 
a publication  on the last day of the window would never obtain citations 
no matter what its eventual impact is. 
However, the normalisation should be 
affected in the same way and thus differences should be small as long as 
the sample is large.  In ADS such a window must be set by hand. 

We determined all these results for a two 3-year periods 
(2000-2002 and 2003-2005)
as well as two 4-year periods (2000-2003 and 2003-2006)
to see whether  such changes in the time 
window are of any influence on the impact ratios.

Detailed results as obtained from the ADS are presented 
in Tables \ref{tabnumNOTW}-\ref{tabnumsamecit} 
at the end of the report.  These tables present the impact ratios for all 
institutes as well as the overall results for the universities and 
for all astronomical institutes. Table \ref{tabnumNOTW} 
shows the results as obtained from the NOWT reports and Tables 
\ref{tabnum4maj}-\ref{tabnumsamecit} the results as obtained from ADS. 
\bigskip

The impact ratios and total numbers of publications
found using the various normalisations described above are
collected in Table \ref {tabnumuni}. It 
shows that, according to the data obtained 
with ADS, the impact of the Dutch articles is {\it increasing} instead of 
decreasing. This trend is seen in all the different methods that were used 
to process the ADS data. This is different from the NOWT results and we will
try to find out why this is so.

First note that for the period 2003-2006 the number of
publications that NOWT uses is already comparable to those in the four
major journals. In total ADS has found many more publications from
authors at the Dutch universities (1721) than NOWT has for all of Dutch
astronomy (1311 or 1407). The situation is is different for the period 
2000-2003, where NOWT and ADS find comparable numbers (1807 versus
1830). In any case, the results for 2003-2006 in the NOWT is derived
from fewer publications than in reality have been published by all Dutch
astronomers. Note that inclusion of the non-academic institutes in the
sample would only have an effect if their workers would pre-dominantly publish
outside the four major journals. However, this is not the case, which
can be seen by comparing the number of 
publications for ASTRON and SRON in Tables \ref{tabnum4maj} and
\ref{tabnumAll}. This shows that also these institutes publish mostly in the
four major journals. 

\hspace*{0.5 cm}Another remarkable feature in Table  \ref {tabnumuni}  is the 
 high impact ratios obtained when considering all refereed journals in 
ADS. This is purely caused by the fact that the bulk of the Dutch  astronomical 
publications are in the major journals, in papers which receive many more 
citations than publications in other journals. Since the 
Dutch astronomical publications mostly appear in the major journals the average 
number of citations per article does not change much when one includes less 
important journals, while the average number of citations per article of 
the field is severely lowered by the inclusion of smaller journals. The CWTS
solves this in their studies by also calculating the impact ratio with respect to
the actual journal citation rates rather than that of the field. This could be 
done in ADS also, but is less straightforwared to implement.

When we consider the number  of publications of NIKHEF we find 
a large difference between NOWT and the number of articles found in ADS. 
The 2008 NOWT states 118 articles whereas ADS produces 11 articles for 
all refereed journals. This is caused by the fact that articles in the journal 
{\it Physical Review D} are always considered to be astrophysical by the ISI 
{\it Web of Knowledge} (WoK) 
whereas ADS splits articles in this journal into astronomy 
articles and physics articles. Interestingly the WoK 
considers 
all these articles also to be physics articles. A quick glance at the articles 
in question shows that the majority of these articles are related to the decay 
or branching of elementary particles. It is questionable whether such articles 
should be considered astrophysical.

ADS finds more publications (in 2003-2006) than NOWT.
We also found that comparing to annual reports even ADS is still missing
publications from the university institutes due to double affiliations.
This should not be a problem when all Dutch insttiutes are considered
since in general such publications then would be found. Still,
there appears a clear underrepresentation of papers in refereed journals
in the NOWT data.  At the same  time all impact scores are higher when
determined with the use of ADS. We believe that the ADS results (both
the values of the impact ratios themselves as the trend with time) are
robust and a better determination of the actual situation than the NOWT
results.

It is also interesting to see that the choice of citations
window does not  affect the trend observed in the impact ratio. However, a 
citation window that coincides with the publication window does significantly 
lower the impact ratio, as would be expected. 

When comparing the number of citations found per article, we 
find that articles in ADS consistently gain 
more citations with respect to the same articles in the WoK. 
This is caused by the fact that the WoK does not count citations 
to the article when the reference is made with a pre-print identifier. In a 
society where the use of  pre-print archives becomes more and more common, 
one can expect this lack of citations in the WoK to become 
increasingly important. Also this effect is most likely more pronounced in 
the NOWT reports because of the short citation windows. 
\bigskip

{\bf Summarizing:}\\
{\it From a citation analysis through the use of ADS we conclude 
that the impact ratio of Dutch astronomical publications is actually rising. 
This trend seems to be firm over several methods of calculating the impact ratio 
and is opposite to what is reported in the NOWT reports. This difference is most 
likely caused by a better separation of astronomy and physics in ADS than in the 
WoK. ADS probably finds more citations in conference 
proceedings, while the inclusion of citations to articles with their 
pre-print identifier could also help explain the difference (especially since 
the citation 
windows in the reports are short). Differences in the actual selection 
process between the different NOWT reports seem to be present and contributing 
to the differences between ADS and the NOWT report.}


\newpage
\section{Figures and tables}

\clearpage

\clearpage

\newpage


\small
\begin{sidewaystable}[htp]
\begin{center}
\begin{tabular}{|ccccccccccccc|}
\hline
\multicolumn{13}{|c|}{Impact ratio}\\
\hline
\hline
 year report (calc)	&Total	&   univ.  &RUG	&UL	&RU	&UvA	&UU	&ASTRON &SRON	&NIKHEF	&Rijnhuizen &KNMI	\\
\hline
2008(2003-2006)		&		&1.19	       &1.19	&1.18	&1.03	&1.26	&1.13	&0.84(s)&1.27(s)&1.13	&0.85	    &0.22	\\
2005(2000-2003)		&1.27		&	       &	&	&	&	&	&	&	&	&	    & 	\\
2003(1998-2001)		&1.29		&1.29	       &1.43	&1.35	&	&1.22	&1.2 &1.66(a)&1.04(a)&	 & &	\\
2000(1994-1998)		&		&1.3	       &	&	&	&	&	&	&	&	&	    &	\\
1998(1992-1996)		&1.07		&1.11	       &	&	&	&	&	&	&	&	&	 &	\\
\hline
\hline
\multicolumn{13}{|c|}{Total number of publications}\\	
\hline
\hline
 year report (calc)	&Total	&   univ.  &RUG	&UL	&RU	&UvA	&UU	&ASTRON &SRON	&NIKHEF	&Rijnhuizen &KNMI	\\
\hline
2008(2003-2006)		&		&1311	       &276	&470	&94	&390	&177	&233(s)	&300(s)	&118	&19	    &10	\\
2005(2000-2003)		&1807		&	       &	&	&	&	&	&	&	&	&	    &	\\
2003(1998-2001)		&		&	       &	&	&	&	&	&114(a)	&463(a)	&	&	    & 	\\
2000(1997-1998)		&		&	       &	&	&	&	&	&	&	&	&	    &	\\
1998(1995-1996)		&390		&	       &	&	&	&	&	&	&	&	&	    &	\\
\hline
\end{tabular}\caption{\small Impact ratios of astronomy from the NOWT reports.}
\end{center}  
(a) All institute publications (technical and astronomical). 
(s) Solely astronomy publications.
\label{tabnumNOTW}
\end{sidewaystable}%

\begin{sidewaystable}[htp]
\centering
\begin{tabular}{|ccccccccccccc|}
\hline
\multicolumn{13}{|c|}{Impact ratio}\\
\hline
\hline
 year report (calc)	&Total	&   univ.  &RUG	&UL	&RU	&UvA	&UU	&ASTRON &SRON	&NIKHEF	&Rijnhuizen &KNMI	\\
2008(2003-2006)		&1.23		&1.27	       &1.28	&1.46	&0.79	&1.20	&1.13	&1.07	&1.14	&0.20	&0.86	    &0.37\\	
2008(2003-2005)		&1.22		&1.27	       &1.23	&1.44	&0.88	&1.20	&1.18	&1.05	&1.14	&0.18	&0.84	    &\\	
2005(2000-2003)		&1.18		&1.21	       &1.17	&1.29	&0.46	&1.25	&1.12	&0.87	&1.22	&	&0.78	    &\\	
2005(2000-2002)		&1.21		&1.22	       &1.14	&1.35	&0.37	&1.26	&1.09	&0.90	&1.30	&	&0.79	    &\\	

\hline
\hline
\multicolumn{13}{|c|}{Total number of publications}\\	
\hline
\hline
 year report (calc)	&Total	&   univ.  &RUG	&UL	&RU	&UvA	&UU	&ASTRON &SRON	&NIKHEF	&Rijnhuizen &KNMI	\\
\hline
2008(2003-2006)		&1812		&1399	       &292	&475	&95	&365	&172	&163	&228	&1	&18	    &3\\	
2008(2003-2005)		&1273		&970	       &212	&325	&52     &264	&117	&119	&167	&1	&16	    &0\\	
2005(2000-2003)		&1570		&1193	       &279	&365	&15	&338	&196	&128	&239	&0	&10	    &0\\	
2005(2000-2002)		&1125		&873	       &211	&260	&6	&245	&151	&80	&165	&0	&7	    &0\\	

\hline
\end{tabular}\caption{\small Impact ratios calculated from ADS solely based on the the 4 
major journals (Astronomy \& Astrophysics, Monthly Notices of the Royal Astronomical 
Society, Astrophysical Journal, Astronomical Journal). Citation window up to december 
2008. Each range of years includes the last year.}
\label{tabnum4maj}
\end{sidewaystable}%

\begin{sidewaystable}[htp]
\centering
\begin{tabular}{|ccccccccccccc|}
\hline
\multicolumn{13}{|c|}{Impact ratio}\\
\hline
\hline
 year report (calc)	&Total	&   univ.  &RUG	&UL	&RU	&UvA	&UU	&ASTRON &SRON	&NIKHEF	&Rijnhuizen &KNMI	\\
\hline
2008(2003-2006)		&1.26		&1.31	       &1.29	&1.47	&0.84	&1.34	&1.10	&1.09	&1.11	&2.07	&0.84	    &0.36\\	
2008(2003-2005)		&1.26		&1.31	       &1.24	&1.46	&0.95	&1.31	&1.15	&1.08	&1.11	&1.86	&0.82	    &\\	
2005(2000-2003)		&1.20		&1.24	       &1.20	&1.30	&0.63	&1.31	&1.13	&0.86	&1.23	&	&0.90	    & \\	
2005(2000-2002)		&1.21		&1.23	       &1.16	&1.34	&0.75	&1.25	&1.11	&0.89	&1.31	&	&0.94	    & \\	

\hline
\hline
\multicolumn{13}{|c|}{Total number of publications}\\	
\hline
\hline
 year report (calc)	&Total	&   univ.  &RUG	&UL	&RU	&UvA	&UU	&ASTRON &SRON	&NIKHEF	&Rijnhuizen &KNMI	\\
\hline
2008(2003-2006)		&1848		&1428	       &296	&483	&98	&379	&172	&169	&228	&2	&18	    &3\\
2008(2003-2005)		&1300		&1182	       &215	&332	&55	&272	&117	&124	&167	&2	&16	    &0\\
2005(2000-2003)		&1603		&1219	       &284	&371	&17	&348	&199	& 131	&242	&0	&11	    &0\\
2005(2000-2002)		&1146		&889	       &214	&263	&8	&250	&154	& 81	&168	&0	&8	    &0\\

\hline
\end{tabular}\caption{\small Impact ratios calculated from ADS solely based on the  
6 major journals (Astronomy \& Astrophysics, Monthly Notices of the Royal 
Astronomical Society, Astrophysical Journal, Astronomical Journal, Science, 
Nature). Citation window up to december 2008. 
Each range of years includes the last year. 
>From Nature and Science only articles listed as astronomical are included.}
\label{tabnum6maj}
\end{sidewaystable}%

\begin{sidewaystable}[htp]
\centering
\begin{tabular}{|ccccccccccccc|}
\hline
\multicolumn{13}{|c|}{Impact ratio}\\
\hline
\hline
 year report (calc)	&Total	&   univ.  &RUG	&UL	&RU	&UvA	&UU	&ASTRON &SRON	&NIKHEF	&Rijnhuizen &KNMI	\\
\hline
2008(2003-2006)		&1.26		&1.31	       &1.29	&1.47	&0.84	&1.34	&1.10	&1.09	&1.11	&2.07	&0.84	    &0.36\\	
2008(2003-2005)		&1.26		&1.31	       &1.24	&1.46	&0.95	&1.31	&1.15	&1.08	&1.11	&1.86	&0.82	    &\\	
2005(2000-2003)		&1.20		&1.24	       &1.20	&1.30	&0.63	&1.31	&1.13	&0.86	&1.23	&	&0.90	    & \\	
2005(2000-2002)		&1.21		&1.23	       &1.16	&1.34	&0.75	&1.25	&1.11	&0.89	&1.31	&	&0.94	    & \\	
\hline
\hline
\multicolumn{13}{|c|}{Total number of publications}\\	
\hline
\hline
 year report (calc)	&Total	&   univ.  &RUG	&UL	&RU	&UvA	&UU	&ASTRON &SRON	&NIKHEF	&Rijnhuizen &KNMI	\\
\hline
2008(2003-2006)		&1931		&1490	       &321	&514	&98	&382	&175	&187	&231	&2	&18	    &3\\
2008(2003-2005)		&1344		&1024	       &228	&349	&55	&274	&118	&134	&168	&2	&16	    &0\\
2005(2000-2003)		&1627		&1238	       &289	&384	&17	&349	&199	&135	&243	&0	&11	    &0\\
2005(2000-2002)		&1150		&893	       &216	&264	&8	&251	&154	&81	&168	&0	&8	    &0\\
\hline
\end{tabular}\caption{\small Impact ratios calculated from ADS solely based on the 6 major 
journals (Astronomy \& Astrophysics, Monthly Notices of the Royal Astronomical 
Society, Astrophysical Journal, Astronomical Journal, Nature, Science) plus two 
smaller journals (Astronomische Nachrichten and New Astronomy Reviews). Citation 
window up to december 2008. Each range of years includes the last year.}
\label{tabnum8maj}
\end{sidewaystable}%
\begin{sidewaystable}[htp]
\centering
\begin{tabular}{|ccccccccccccc|}
\hline
\multicolumn{13}{|c|}{Impact ratio}\\
\hline
\hline
 year report (calc)	&Total	&   univ.  &RUG	&UL	&RU	&UvA	&UU	&ASTRON &SRON	&NIKHEF	&Rijnhuizen &KNMI	\\
\hline
2008(2003-2006)		&1.89		&2.00	       &1.99	&2.19	&1.32	&2.11	&1.70	& 2.45	&1.67	&1.29	&0.93	    &0.34\\	
2008(2003-2005)		&1.90		&2.00	       &1.92	&2.17	&1.56	&2.07	&1.71	&2.54 	&1.62	&1.29	&0.93	    &0.20\\	
2005(2000-2003)		&1.85		&1.89	       &1.83	&1.91	&1.02	&2.08	&1.76	&1.90	&1.82	&0.39	&1.36	    &0.14 \\	
2005(2000-2002)		&1.81		&1.87	       &1.76	&2.01	&1.12	&1.96	&1.71	&1.32	&1.93	&0.27	&1.53	    &0.13\\	
\hline
\hline
\multicolumn{13}{|c|}{Total number of publications}\\	
\hline
\hline
 year report (calc)	&Total	&   univ.  &RUG	&UL	&RU	&UvA	&UU	&ASTRON &SRON	&NIKHEF	&Rijnhuizen &KNMI	\\
\hline
2008(2003-2006)		&2261		&1721	       &349	&593	&113	&430	&236	&215	&273	& 11	&31	    &10\\
2008(2003-2005)		&1578		&1182	       &252	&406	&60	&308	&156	&152	&203	&8	&26	    &7\\
2005(2000-2003)		&1830		&1376	       &315	&438	&20	&371	&232	&155	&280	&5	&12	    &2\\
2005(2000-2002)		&1285		&986	       &233	&298	&11	&266	&178	&91	&195	&3	&8	    &2\\
\hline
\end{tabular}\caption{\small Impact ratios calculated from ADS based on all refereed 
journals as listed by ADS. Citation window up to december 2008. Each range of years 
includes the last year.}
\label{tabnumAll}
\end{sidewaystable}%
\begin{sidewaystable}[htp]
\centering
\begin{tabular}{|ccccccccccccc|}
\hline
\multicolumn{13}{|c|}{Impact ratio}\\
\hline
\hline
 year report (calc)	&Total	&   univ.  &RUG	&UL	&RU	&UvA	&UU	&ASTRON &SRON	&NIKHEF	&Rijnhuizen &KNMI	\\
\hline
2008(2003-2006)		&1.80		&1.87	       &1.83	&1.99	&1.15	&2.09	&1.56	&1.60	&1.61	&1.48	&0.99	    &0.17\\	
2008(2003-2005)		&1.86		&1.93	       &1.87	&1.97	&1.18	&2.17	&1.74	&1.69 	&1.74	&1.60	&0.92	    &0.08\\	
2005(2000-2003)		&1.71		&1.78	       &1.72	&1.76	&0.92	&1.90	&1.77	&1.08	&1.74	&0.26	&1.20	    &0.32 \\	
2005(2000-2002)		&1.74		&1.81	       &1.77	&1.91	&1.33	&1.89	&1.64	&1.23	&1.64	&0.37	&1.37	    &0.14\\	

\hline
\hline
\multicolumn{13}{|c|}{Total number of publications}\\	
\hline
\hline
 year report (calc)	&Total	&   univ.  &RUG	&UL	&RU	&UvA	&UU	&ASTRON &SRON	&NIKHEF	&Rijnhuizen &KNMI	\\
\hline
2008(2003-2006)		&2261		&1721	       &349	&593	&113	&430	&236	&215	&273	& 11	&31	    &10\\
2008(2003-2005)		&1578		&1182	       &252	&406	&60	&308	&156	&152	&203	&8	&26	    &7\\
2005(2000-2003)		&1830		&1376	       &315	&438	&20	&371  &232	&155	&280	&5	&12	    &2\\
2005(2000-2002)		&1285		&986	       &233	&298	&11	&266	&178	&91	&195	&3	&8	    &2\\

\hline
\end{tabular}\caption{\small  Impact ratios calculated from ADS based on all refereed 
journals as listed by ADS . Citation window is equal to the publication period. 
Each range of years includes the last year.}
\label{tabnumsamecit}
\end{sidewaystable}%

\clearpage
\newpage

\section{Appendix A}
{\normalsize
The 15 instutes that were considered  as top in US are listed below. This list
was constructed by taking the astronomy departments of the first 13 instutes as listed in Table 1 of  A.L. Kinney's  ``The
Science Impact of Astronomy PhD Granting Departments in the United
States''\footnote{arXiv:0811.0311} and supplemented with UCLA and the
University of Texas at Austin. For our analysis only faculty that is listed as
active and part of astronomy are considered. Faculty that is listed as
physicist is excluded from our sample.

1 Caltech
 
2 UC Santa Cruz 

3 Princeton University

4 Harvard University 

5 Colorado

6 SUNY Stony Brook 

7 Johns Hopkins University

8 Penn. State Univ.
 
9 Univ. Michigan

10 Univ. Hawaii 

11 Univ. Wisconsin

12 UC Berkeley 

13 Michigan State Univ.
 
14 UCLA

15 Univ. Texas
}
\end{document}